\documentstyle[12pt]{article}
\input{epsf}
\def \inbar{\vrule height1.5ex width.4pt depth0pt}
\def \C{\relax\hbox{\kern.25em$\inbar\kern-.3em{\rm C}$}}
\def \R{\relax{\rm I\kern-.18em R}}

\newcommand{\beq}{\begin{equation}}
\newcommand{\eeq}{\end{equation}}
\newcommand{\bea}{\begin{eqnarray}}
\newcommand{\eea}{\end{eqnarray}}
\newcommand{\nn}{\nonumber}

\newcommand{\pdr}{\partial} 
\newcommand{\tE}{\tilde E}

\newcommand{\eps}{\epsilon}
\newcommand{\tr}{\hbox{tr}}
\newcommand{\Tr}{\hbox{Tr}}

\newcommand{\Hi}{\cal H}
\begin{document}

\title{Geometric Quantization and  Two  Dimensional QCD}
\author{S. G. Rajeev\thanks{email:rajeev@urhep.pas.rochester.edu} and  O. T.
Turgut\thanks{  email:turgut@ihes.fr}\\  
     \\
  University of Rochester\\
   Rochester, NY, 14623, USA\\
   and\\
   IHES, 35 Route de Chartres,\\
    91440, Bures-sur-Yvette, France.}

\date{May 1997}
\maketitle
\begin{abstract}
In this article, we will discuss geometric quantization of 2d QCD
with fermionic and bosonic matter fields.
We  identify the respective large-$N_c$ 
phase spaces as the infinite dimensional 
Grassmannian and the infinite dimensional Disc. The Hamiltonians 
are quadratic functions, and 
the resulting equations of motion for these classical systems are nonlinear.
In \cite{2dqhd}, it was shown that the linearization of the 
equations of motion for the Grassmannian gave the `t Hooft equation.
We will see that the linearization in the bosonic case leads to 
the scalar analog of the `t Hooft equation found in \cite{Tomaras}. 
\end{abstract}

\section{ Introduction}

In the large $N_c$ limit of various quantum field  theories  (e.g., Quantum
Chromodynamics or QCD) the quantum fluctuations become small and the theories
tend to a classical limit. This classical limit however is different from the
conventional one, in that many of the essential non--perturbative features of
the quantum theory survive the large $N_c$ limit\cite{thooft2, tHooft, 
witten2}.  These classical theories
are somewhat unusual in that they are not local field theories. Moreover the
phase spaces are some curved manifolds, and some  of the nonlinearities
(interactions) of the theory arises from this curvature. A point of view to
these theories starting from the geometry of these phase spaces is therefore
natural. Pioneering work in this direction was done by Berezin \cite{berezin}.
 The first author 
 has revived this point of view more recently \cite{2dqhd}. ( See also
\cite{wadia,mandal,cavicchi,kikkawa,kikkawa2}.) 

The situation is analogous to that in atomic physics. In the conventional
classical limit the atom is unstable due to radiation from the orbiting
electrons. However, in the mean field (or Hartree--Fock ) approximation, the
atom is stable. In this  approximation quantum  fluctuations in the density
matrix of electrons is ignored. What is perhaps less well appreciated is that
the Hartree--Fock approximation of atomic physics is equivalent to a classical
theory whose phase space is an infinite dimensional complex manifold
\cite{2dqhd}.
(Indeed this approximation can be thought of as a large $N_c$ limit of the
atomic system.) The density matrix itself is treated as  a dynamical variable:
the set of values it can take is the phase space of the theory.

We will develop the geometric point of view to these theories
further, by treating fermionic and bosonic systems in parallel. It turns out
that many 
of the mathematical ideas are simpler  for bosons, since the phase space
admits a global co--ordinate system.
In the case of fermionic systems,
the appropriate phase spaces are coset spaces of unitary groups called the
Grassmannian. These are well known objects in algebraic geometry in the finite
dimensional case \cite{chern,griffiths}. 
The corresponding infinite dimensional
generalization is essentially due to G. Segal \cite{pres-segal, segal}.
  We will show that
the phase space of the large $N_c$ limit of the bosonic systems is a coset
space of the pseudo--unitary group. This  space is a non--commutative analog
of the open disc in the complex plane: it is the space of matrices $Z$ such
that $Z^{\dagger}Z<1$.  The pseudo-unitary group acts by fractional linear
transformations on this space. We will find also the appropriate infinite
dimensional version of this space relevant to two dimensional field theory.

First we will describe a classical dynamics on  these coset spaces. 
We will then quantize these classical systems using
geometrical ideas. 
The special case of the large $N_c$ limit of scalar QCD in two space--time
dimensions will be discussed in more detail. The linear approximation to this
theory can then be obtained. We will see that this agrees with the traditional
large $N_c$ analysis, which in this case was carried out by Tomaras 
\cite{Tomaras}
following the original ideas of 't Hooft \cite{tHooft}.

It is important to emphasize that the large $N_c$ limit of QCD we obtain is a
highly nonlinear classical theory. It is only the linear approximation that can
be obtained by conventional methods of summing planar diagrams. The geometrical
approach encodes the nonlinearities in the curvature of the underlying phase
space.

\section{The Disc and the  Grassmannian}

We will describe first the geometry of two homogenous symplectic  manifolds, the  
Disc and the Grassmannian. The quantization of the dynamical systems with these 
as phase spaces lead to a bosonic and  a  fermionic  system respectively.
If the dimension of the homogenous spaces are finite, this will be a quantum 
mechanical system with a finite number of degrees of freedom. But our discussion 
will 
  be tailored to apply to infinite dimensional cases subject to some convergence
   conditions; 
these corresponds to bosonic or fermionic quantum field theory in two 
space--time 
dimensions. It is at the moment not known how to generalize this discussion to 
the  (less convergent) infinite dimensional manifolds    that  correspond to 
quantum field theories in higher dimensional space--times. Our approach is much 
influenced 
by the discussion of the Grassmannian in the book by Pressley and Segal 
\cite{pres-segal}.

Let $\Hi$ be a separable complex Hilbert space; $\Hi_-$ and $\Hi_+$ are two 
orthogonal subspaces with $\Hi=\Hi_-\oplus \Hi_+$.  Define the Disc 
$D_1(\Hi_-,\Hi_+)$ to be the set of all operators $Z:\Hi_+\to \Hi_-$ such that 
$1-Z^{\dag}Z>0$ and $Z$ is Hilbert-Schmidt: $\tr Z^{\dag}Z<\infty$.      The 
last 
condition is automatically satisfied if either $\Hi_-$ or $\Hi_+$  is finite 
dimensional. 

Next, we define the  pseudo-unitary group to be
a subset of the invertible operators
from ${\cal H}$ to $\cal H$:
\beq
       U_1({\cal H}_-,{\cal H}_+)
             =\{ g| g\epsilon g^{\dagger}=\epsilon, \quad g^{-1}\ {\rm
                    exists\ \ and } \ [\epsilon, g] {\rm\  is\ \  
Hilbert-Schmidt} 
\}
.\eeq
Again, if the dimension of $\Hi_+$ or $\Hi_-$ is  finite, the last condition is 
unnecessary. 
Here $\eps=\pmatrix{-1&0\cr 0&1}$ with respect to the decomposition 
$\Hi=\Hi_-\oplus \Hi_+$.
If we decompose the matrix into block forms,
\beq
        g=\pmatrix{ a& b\cr c&d\cr}  \nn
\eeq
we have, $a:{\cal H}_- \to {\cal H}_-$, $b:{\cal H}_+ \to {\cal H}_-$
, $c:{\cal H}_- \to {\cal H}_+$ and $d:{\cal H}_+ \to {\cal H}_+$.  Then, the  
off diagonal elements $b$ and $c$
 are Hilbert-Schmidt  and the diagonal elements $a$ and $d$
are bounded and invertible operators. The space of 
Hilbert-Schmidt operators form a two-sided ideal (which we will denote by ${\cal 
I}_2$) in the algebra of bounded operators. Thus the condition on the 
off-diagonal 
elements is preserved by multiplication and inversion, so that $U_1$ is indeed a 
group. The geometric meaning of the condition on $[\eps,g]$ is that the linear 
transformation $g$ does not mix the subspaces $\Hi_\pm$ by `too much'.

We define an action of $U_1({\cal H}_-,{\cal H}_+)$ on the Disc $D_1$:
\beq
     Z \mapsto g \circ Z=(aZ+b)(cZ+d)^{-1}.
\eeq
 The condition $1-Z^{\dag}Z>0$ implies that $cZ+d$ is invertible and bounded. 
Since the space of Hilbert--Schmidt operators is a two-sided ideal, 
$(aZ+b)(cZ+d)^{-1}$  is still Hilbert-Schmidt. Thus our  action is 
well--defined.
 
 We  note that the  stability
subgroup of the point $Z=0$ is $ U({\cal H}_-) \times U({\cal H}_+)$, 
$U(\Hi_\pm)$
being the group  of {\it all} unitary  operators on $\Hi_\pm$.
Moreover, any point  $Z$ is the image of $0$ under the action of the group,
$g \circ (Z=0)=bd^{-1}$. (note that $bd^{-1}$ is in ${\cal I}_2$ and
$d^{\dagger}d=1+b^{\dagger}b$ implies that $1-(bd^{-1})^{\dagger}bd^{-1}>0$).
We therefore see that $D_1$ is a homogeneous space and
given by the quotient;
\beq
      D_1={ U_1({\cal H}_-,{\cal H}_+) \over
            U({\cal H}_-) \times  U({\cal H}_+)}
.\eeq

 It will prove  convenient to parametrize the Disc by operators  $\Phi:{\cal H} 
\to {\cal H}$,
\beq
     \Phi=1-2\pmatrix{ (1-ZZ^{\dagger})^{-1}
                                 &-(1-ZZ^{\dagger})^{-1}Z\cr
                         Z^{\dagger} (1-ZZ^{\dagger})^{-1}
                      &-Z^{\dagger}(1-ZZ^{\dagger})^{-1}Z\cr}
.\eeq
One can see that under the transformation $Z\mapsto 
 g\circ Z$, $\Phi \mapsto g^{-1}\Phi 
g$. $\Phi$
satisfies $\epsilon \Phi^{\dagger} \epsilon=\Phi$ and
$\Phi^2=1$. Also, $\Phi-\epsilon \in {\cal I}_2$, so 
that as an operator $\Phi$ does not differ from $\epsilon$ by
 `too much'. Many physical quantities are best described as functions of 
the  deviation of $\Phi$  from 
the standard value $\eps$, $M=\Phi-\eps$.  We will see that $\eps$ corresponds 
to the 
vacuum state, so this vacuum subtraction is the geometric analogue of normal 
ordering in 
quantum field theory. On occassion, we may use the 
self-adjoint  operator $\tilde M=M\eps$.

Given a complex Hilbert space $\Hi=\Hi_-\oplus \Hi_+$ 
  and orthogonal subspaces $\Hi_\pm$ as before, we can define another homogenous 
space, the Grassmannian.  
We define the Grassmannian to be the  following  set of operators on $\cal H$:
\beq
     Gr_1=\{\Phi|\Phi=\Phi^{\dag};\Phi^2=1;\Phi-\epsilon\in
{\cal I}_2\}.
\eeq
This  is  the same
as the restricted Grassmannian of Ref.\cite{pres-segal} if $\Hi_\pm$ are 
infinite 
dimensional.

To each point in the Grassmannian there corresponds a subspace of $\Hi$, the 
eigenspace of $\Phi$ with eigenvalue $-1$. In fact the  Grassmannian is viewed 
usually as the set of subspaces of a Hilbert space. If $\Hi$ is finite 
dimensional,
the Grassmannian is a compact manifold. It is disconnected, with each connected 
component labelled by $\tr{1-\Phi\over 2}=0,1,\cdots \dim \Hi$.

$ Gr_1$ is the homogeneous space of a 
unitary group. If $\Hi_\pm$ are infinite dimensional, in order to have a 
well--defined
action on $Gr_1$, we must restrict to an appropriate sub--group of
$ U(\cal H)$. We define,
\beq
      U_1({\cal H})=\{g|g^{\dag}g=1;[\eps,g]\in {\cal I}_2\}.
\eeq
Let  us split $g$ into $2\times 2$ blocks
\beq
     g=\pmatrix{g_{11}&g_{12}\cr g_{21}&g_{22}}. \label{sp1}
\eeq
 The convergence condition on $[\eps,g]$ is the  statement
that the off--diagonal blocks $g_{12}$ and $g_{21}$ are in
${\cal I}_2$. It then follows, in the  case where
 $\Hi_\pm$ are both infinite dimensional, that $g_{11}$ and $g_{22}$ are
Fredholm operators.(To see this, we recall that an operator
is Fredholm if it is invertible modulo a compact operator.
Any operator in ${\cal I}_2$ is  compact and moreover,  $g$ is
invertible.) The Fredholm index of $g_{11}$ is opposite to
that of $g_{22}$; this integer is a homotopy invariant of
$g$ and we can decompose $U_1(\cal H)$ into connected components
labeled by this integer.

With the projection $g\to g\eps g^{\dag}$, we see that
$ Gr_1$ is a homogeneous space of $U_1(\cal H)$:
\beq
      Gr_1= U_1({\cal H})/U({\cal H}_-)\times  U({\cal H}_+).
\eeq
For, any $\Phi\in Gr_1$ can be diagonalized by an element of
$ U_1({\cal H})$, $\Phi=g\eps g^{\dag}$; this $g$ is ambiguous up to
right multiplication by an element that commutes with
$\eps$. Such elements form the subgroup
\beq
      U({\cal H}_-)\times  U({\cal H}_+)=\{h|h=\pmatrix{h_{11}&0\cr
0&h_{22}}; h_{11}^{\dag}h_{11}=1=h_{22}^{\dag}h_{22}\}.
\eeq
Each point $\Phi\in  Gr_1$ corresponds to a subspace of
$\cal H$: the eigenspace of $\Phi$ with eigenvalue $-1$. Thus
$Gr_1$ consists of all subspaces obtained from ${\cal H}_-$ by an
action of $U_1$.

It is also possible to view $Gr_1$ as a coset space of
complex Lie groups: this defines a complex structure on
$Gr_1$ which will be useful for geometric quantization.
Define the restricted general linear group
\beq
     GL_1=\{\gamma|\gamma \ \hbox{\rm  is  invertible};
[\eps,\gamma]\in {\cal I}_2\}.
\eeq
Again if we were to decompose into $2\times 2$ submatrices
$\gamma_{12},\gamma_{21}\in {\cal I}_2$ while $\gamma_{11}$ and $\gamma_{22}$
are Fredholm.
Define also the  subgroup ( `Borel subgroup') of matrices
which are upper triangular in this decomposition,
\beq
     B_1=\{\beta=\pmatrix{\beta_{11}&\beta_{12}\cr
0&\beta_{22}}|\beta\in GL_1\}.
\eeq
This is the stability group of ${\cal H}_-$ under the action of
$GL_1$ on ${\cal H}$. Thus the Grassmannian (which is the orbit
of ${\cal H}_-$) is the complex coset space,
\beq
     Gr_1=GL_1/B_1.
\eeq
The tangent space to the Grassmannian at $\eps$ may be
identified with the Hilbert space ${\cal I}_2({\cal H}_-; {\cal H}_+)$.

In finite dimensions  the  Grassmannian and Disc have  a
symplectic structure $\omega= {i\over 4}\tr \Phi d\Phi
d\Phi$. It is obvious that this two--form is invariant under the action of the 
unitary group on the Grassmannian  and the pseudo--unitary group on the Disc,
 since they can both be written as 
,$\Phi\mapsto g\Phi g^{-1}$.  Now, recall that in both cases $\Phi^2=1$ so that 
$\Phi d\Phi=-d\Phi\Phi$. Hence, 
$d\omega=\tr(d\Phi)^3=\tr\Phi^2(d\Phi)^3=-\tr\Phi(d\Phi)^3\Phi=-\tr(d\Phi)^3\Phi
^
2=-d\omega$ which proves that $\omega$ is closed. Due to the homogenity, it is 
enough to prove that $\omega$ is non--degenerate at one point, say $\Phi=\eps$, 
which is also straightforward.

It is not clear that this symplectic form  exists in the
infinite dimensional case; the trace could  diverge.
But if we think in terms of  the
variable $M$,  obtained from $\Phi$ through a `vacuum
subtraction',
 we see  that
\beq
     \omega= {i\over 4}\Tr(\eps+M)dMdM.
\eeq
 is well-defined because $dM \in {\cal I}_2$. 
 Indeed this is why we imposed the convergence conditions. It 
is possible to weaken  the convergence condition ( which is interesting for 
quantum field theories in dimensions greater than two \cite{mickellson}),  
without changing much of the structure but we will lose the symplectic form.

The above form is invariant under the action of $U_1(\cal H)$ for the
Grassmannian and invariant under the action of $U_1({\cal H}_-,{\cal H}_+)$
for the Disc. 
Thus,  $Gr_1$  and $D_1$ are both
homogeneous symplectic manifolds just as in the finite
dimensional case. We can look for
the moment maps, which generate the infinitesimal action of
$U_1({\cal H}_-,{\cal H}_+)$ and $U_1({\cal H})$ respectively.
In the finite dimensional case, this is just the function $-\Tr u\Phi$, where 
$u$ 
is a hermitian matrix for the Grassmannian and a pseudo hermitian 
($u^{\dag}=\eps u\eps$) matrix for the Disc. Indeed, the infinitesimal change of 
$\Phi$ under the group is $[u,\Phi]$, and $2\Tr\Phi[u,\Phi]d\Phi=-d\Tr u\Phi$.

We cannot take    $f_u=-\Tr u \Phi  $ in the infinite dimensional case,
because the trace  diverges. However,  we do a vacuum subtraction
from this expression and get instead $-\Tr (\Phi-\epsilon)u=-\Tr Mu$; this
trace is
 conditionally convergent, so we have a chance of obtaining a generating 
function. 
 
 We will now describe this procedure in more
detail.
Let us  decompose the operator $M$ into block form,
\beq
      M=\pmatrix{M_{11}&M_{12}\cr M_{21}&M_{22}\cr}
\eeq
where, $M_{11}:{\cal H}_- \to {\cal H}_-$, $M_{12}:{\cal H}_+ \to {\cal H}_-$,
 $M_{21}:{\cal H}_- \to {\cal H}_+$, and $M_{22}:{\cal H}_+ \to {\cal H}_+$;
since $\Phi-\epsilon$ is in ${\cal I}_2$ in both cases,
we have $M_{12}, M_{21} \in {\cal I}_2$. If we use the quadratic
constraint, $M^2+[\epsilon, M]_+=0$,
\beq
     M_{11}^2+M_{12}M_{21}-2M_{11}=0,\quad
M_{22}^2+M_{21}M_{12}+2M_{22}=0.
\eeq
Next we use  the fact that $M_{12}, M_{21} \in {\cal I}_2$ and the above
equations to  get
 $M_{11}, M_{22} \in {\cal I}_1$.

 Now, $u \in\pmatrix{{\cal B}&{\cal I}_2\cr {\cal I}_2&{\cal B}}$,
and $M\in\pmatrix{{\cal I}_1&{\cal I}_2\cr {\cal I}_2&{\cal I}_1}$. (Here
 ${\cal B}$ is the space of bounded operators.) Thus the diagonal blocks in
$Mu$ are both trace--class.
We now  define the conditional trace $\Tr_\eps$ of an operator
 to be the sum of the traces of its diagonal submatrices:
$\Tr_{\eps} X=\frac{1}{2}\Tr[X+\eps X\eps]$.
(Such conditional traces have been used recently to study
anomalies \cite{langmann}.) This conditional trace exists for $Mu$ and we define
\beq
	f_u=-\Tr_\eps Mu.
\eeq

If we restrict to finite rank matrices $u$, this function
differs by a constant from the previous moment map;
therefore it generates the same Hamiltonian vector fields:
\beq
     \omega(V_{f_u},.)=-df_u\mapsto V_{f_u}=i[u,\eps+M].
\eeq
However, there is an important change in the Poisson bracket
relations; they will differ by a constant term from the
previous ones:
\beq
     \{f_u,f_v\}=f_{-i[u,v]}-i\Tr_\eps \eps[u,v].
\eeq
In the finite dimensional case we can remove the extra term by adding a
constant term to   $f_u$. However this is not possible in the infinite
dimensional case, as the term we must add to $f_u$ will diverge.
This is, in fact,  the Lie algebra of the non--trivial central extension of
$GL_1$.

It will be convenient to identify the Hilbert space ${\cal H}$ with
$L^2({\bf R})$ of square integrable functions on the real line.
(Since all abstract infinite dimensional  Hilbert spaces are isomorphic, in fact
there is no loss of generality. But this realization is  convenient for 
application to two--dimensional quantum field theories.)
The operator $\eps$ can be thought of as an integral operator with kernel
$\eps(p,q)$.   We can choose
$\eps(p,q)={\rm sgn}(p)\delta(p-q)$.
A natural choice
of basis for the operators on this space is the set of  the Weyl operators
$e(p,q)$  defined by the integral kernels $e(p,q)(r,s)=\delta(p-s)\delta(q-r)$.
Define also
$\lambda(p,q)=\int\epsilon(p,r)e(r,q) [dr]$.

We can express the Poisson brackets   in terms of the expansions
 $M=\int M(p,q)e(q,p)[dpdq]$
for the Grassmannian and $M=\int \tilde M(p,q)\lambda(q,p)[dpdq]$ for the
Disc\footnote{to be more precise one should use a countable basis. It is 
possible to 
consider the functions on the circle and let the radius go to infinity.}:
\bea
   &\ &\{M(p,q),M(r,s)\}=\nn\\
   &\ &i(M(r,q)\delta(p-s)-M(p,s)\delta(r-q))
     +i(\epsilon(r,q)\delta(p-s)-\epsilon(p,s)\delta(r-q)).
\eea
and
\bea
   &\ &\{\tilde M(p,q),\tilde M(r,s)\}=\nn\\
   &\ &i(\tilde M(r,q)\epsilon(p,s)-
       \tilde M(p,s)\epsilon(r,q))
     +i(\epsilon(p,s)\delta(r-q)-\epsilon(r,q)\delta(p-s)).
\eea

\section{Algebraic Quantization}

So far we have discussed the geometrical structures associated to classical 
physics, symplectic geometry. The passage to the quantum theory can be made most 
directly by an algebraic method: find an irreducible representation for  the 
above commutaion reltions of the moment maps. In the infinite dimensional case, 
the central extension will play an important role in this story.

Let us start with operators satisfying  fermionic anticommutation relations:
\beq
    [\chi^\alpha(p),\chi^{\dagger}_\beta(q)]_+=\delta(p-q)\delta^\alpha_\beta;
\eeq
 all the other anticommutators vanish. (We have introduced an extra label 
(`color') $\alpha$ taking values $1,\cdots N_c$ in order to get the most general 
interesting representations of the algebra.)
Here,  $\chi^\alpha(p)$ and  $\chi^{\dagger}_\alpha(p)$ are to be thought of  as 
operator
valued
distributions.
The fermionic Fock space $\cal F$ can be built from the vacuum state $|0>$
, defined by $\chi^\alpha(p)|0>=0$ for $p\ge 0$ and
$\chi^{\dagger}_\alpha(p)|0>=0$
for $p<0$,
by the action of $\chi^\alpha(p)$'s and $\chi^{\dagger}_\alpha(q)$'s.
We define the hermitian operators acting on the Fock space $\cal F$;
$\hat M(p,q)= -{2 \over
N_c}\sum_\alpha:~\chi^{\dagger}_\alpha(p)\chi^\alpha(q)~:$
. As is common in quantum field theory, we define the normal ordering of pairs 
of 
fermionic operators by;
\beq
      :\chi^{\dagger}_\alpha(p)\chi^\alpha(q):=
    \cases{\chi^{\dagger}_\alpha(p)\chi^\alpha(q) & if $p \ge 0$\cr
           -\chi^\alpha(q)\chi_\alpha^{\dagger}(p)& if $p < 0$\cr}
.\eeq

Let us pause to explain why such an ordering rule is necessary. In order to 
represent the Lie algebra of the restricted unitary group, $\int \hat 
M(p,q)u(p,q)[dpdq]$ must be well--defined on the Fock space, where 
$u\in\pmatrix{{\cal B}&{\cal I}_2\cr {\cal I}_2&{\cal B} }$. Now the vacuum 
expectation value $<0|\int \hat M(p,q)u(p,q)[dpdq]|0>$ would have diverged if we 
had not used the normal ordered product. With the normal ordered product, this 
expectation value vanishes. Moreover, the norm of the state $||\int \hat 
M(p,q)u(p,q)[dpdq]|0>||^2$ is finite since only the off--diagonal, 
Hilbert--Schmidt,   part of $u$ contributes to it. This is indeed why we imposed  
the convergence conditions in the definition of the restricted unitary group.

After some algebra one can check that
\bea
  &\ &[\hat M(p,q),\hat M(r,s)]=\nn\\
   &\ &{2 \over N_c}\{\hat M(r,q)\delta(p-s)
    -\hat M(p,s)\delta(r-q)+({\rm sgn}(r)-{\rm sgn}(s))\delta(r-q)\delta(p-s)\}
.\eea
We see that if we identify $\epsilon(r,q)=\delta(r-q){\rm sgn}(r)$ and
$\hbar={2 \over N_c}$ we obtain
a unitary representation of the central extension of $U_1({\cal H})$
 in the fermionic
Fock space.(The same central 
extension was obtained in the mathematics literature by Kac and Peterson
 \cite{kac}).

We introduce,  in a similar fashion, operators satifying the bosonic commutation 
relations.
\beq
     [a^\beta(p),a^{\dagger}_\alpha(s)]={\rm sgn}(p)
                 \delta(p-s)\delta^\beta_\alpha;
\eeq
where, $\alpha,\beta=1,..N_c$ and all the other commutators vanish. The
bosonic  Fock space  is constructed from a vacuum state, defined
through
$a^\alpha(p)|0>=0$ for $p \ge 0$ and $a^\dagger_\alpha(p)|0>=0$ for $p<0$,
 applying the operators $a^\alpha(p)$ and $a^\dagger_\alpha(p)$.
We introduce operators;
${\hat {\tilde M}}(p,q)
={2 \over N_c}\sum_\alpha:a^{\dagger}_\alpha(p)a^\alpha(q):$
with the normal ordering prescription given by,
\beq
     :a^{\dagger}_\alpha(p)a^\alpha(q):=
     \cases{ a^{\dagger}_\alpha(p)a^\alpha(q)& if $p \ge 0$\cr
             a^\alpha(q)a^{\dagger}_\alpha(p)& if $ p<0$.\cr}
\eeq
We can calculate the commutation relations;
\bea
     &\ &[{\hat {\tilde M}}(p,q), {\hat {\tilde M}}(r,s)]=\nn\\
   &\  &{2 \over N_c} \{\hat {\tilde M}(r,q)\epsilon(p,s)-
       \hat {\tilde M}(p,s)\epsilon(r,q)
     +\epsilon(p,s)\delta(r-q)-\epsilon(r,q)\delta(p-s)\}.
\eea
where we identified  $\epsilon(p,s)=-{\rm sgn}(p)\delta(p-s)$.
Note that here $\hat {\tilde M}(p,q)^{\dagger}=\hat {\tilde M}(q,p)$.
This shows that there
is a quantization  of the Poisson Brackets
on the bosonic Hilbert space.

As they stand, these will not be irreducible
representations: the particle number $\hat N$ 
and the `color' operators $\hat Q^\alpha_\beta$, which
generate a $SU(N_c)$ symmetry,  commute
with them.
We can check that for the fermionic operators,
$\hat N=\int~:~\chi^\dagger_\alpha(p)\chi^\alpha(p)~:[dp]$  and
$\hat Q^\alpha_\beta=\int (:~\chi^\dagger_\beta(p) \chi^\alpha(p)~:
-{1 \over N_c}
\delta^\alpha_\beta:~\chi^\dagger_\gamma(p) \chi_\gamma(p)~:)[dp]$.
To obtain an irreducible representation, we need to fix the color
number to zero and the Fermion number to a fixed finite value.
Similarly, in the bosonic Fock space
$\int :~a^\dagger_\alpha(p)a^\alpha(q)~:\epsilon(q,p)[dpdq]$ has to 
be fixed to a
finite value and the color operator
$\hat Q^\alpha_\beta=\int(:~a^\dagger_\beta(p)a^\alpha(q)~:-{1 \over N_c}
\delta^\alpha_\beta:~a^\dagger_\gamma(p)a^\gamma(q)~:)\epsilon(q,p)[dpdq]$
will be put
equal to zero.

\section{Geometric Quantization}

Although it is possible to use this algebraic method to complete the 
quantization 
of the system, many ideas are much clearer in the geometric method. 

We will start by describing   classical mechanics in geometric
terms\cite{arnold,abraham}. 
Let us
assume that
the phase space,
$\Gamma$ is a smooth manifold on which a closed and non-degenerate
2-form $\omega$ is defined.
The observables of a classical system are smooth functions on the phase space.
Since $\omega$ is non-degenerate, given a smooth function
$f$ on $\Gamma$ we can find a vector field generated by $f$;
\beq
        -df=i_{V_f}\omega
.\eeq
This allows us to define the Poisson brackets of two functions $f_1$ and $f_2$;
\beq
       \{ f_1,f_2 \}=\omega(V_{f_1},V_{f_2})
.\eeq
Time evolution for the observables is defined if we choose a Hamiltonian
function $E$;
\beq
       {df  \over dt}=\{f, E\}
.\eeq

Quantization of a given classical system
requires finding an appropriate Hilbert space such that the functions
on phase space are replaced by self-adjoint
operators acting on this Hilbert space. 
There is no unique prescription
for quantization. The first step in geometric quantization is to
find a pre-quantum Hilbert space \cite{axelrod,hurt,kirillov2}.
 If the symplectic form $-i\omega \over \hbar$
belongs to the integral cohomology $H^2(\Gamma,{\bf Z})$, one can construct a
line bundle on $\Gamma$ with Chern class $\left[{-i\omega\over \hbar}\right]$.
There is then a connection on this
line bundle with curvature given by $-i\omega \over \hbar$ \cite{wells,milnor}.
Here $\hbar$ denotes the quantization parameter, $\hbar\to 0$ being the
classical limit.
Square integrable sections of this line bundle provides the `prequantum Hilbert
space' ${\cal H}_{Pre}$.
If we denote the  correspondence between  real functions and self-adjoint
operators acting on this Hilbert space via
  $f \mapsto \tilde f$, then we require
\beq
     \widetilde{ \{ f_1,f_2 \} }={i\over \hbar} [\tilde f_1, \tilde 
f_2]\label{qua}
\eeq
It is possible to realize the above representation of the Poisson
algebra by  using the connection on ${\cal H}_{Pre}$,
 and the answer is given by,
\beq
    \tilde f=-i\hbar \nabla_{V_f}+f
.\eeq
where $V_f$ denotes the vector field generated by $f$.

The mathematical disadvantage  of  prequantization is that this
representation of the Poisson algebra  is highly reducible. Intuitively
the  wave
functions depend on both momenta and coordinates,
which is incompatible with the uncertainty principle. To remedy this we
can restrict to a subspace of sections which are independent of
 `half' of the degrees of freedom. This procedure is called
picking a polarization. In this article our applications will be on Homogeneous
K\"ahler spaces for which there is a natural choice of polarization, the
complex polarization;
thus we will
only talk about this  special case.
(Quantization on K\"ahler manifolds was first considered in detail by 
Berezin, using symbols of operators \cite{berezin2,berezin3,berezin4}).  
We require $\nabla_{\bar i} \psi=0$ on
sections as our choice of polarization,
where ${\bar i}$ denotes the antiholomorphic coordinates.
The integrability condition for  holomorphic sections of line bundles is given
by,
\beq
      [ \nabla_{\bar i}, \nabla_{\bar j} ]\psi=0
.\eeq
One can split the tangent and cotangent
spaces into tensors of type-(r,s) according to the occurance of holomorphic
and antiholomorphic components \cite{kobayashi}. This expression is  equal to
 (0,2) component of the curvature and in our case is proportional to
$\omega$. If the symplectic form  $\omega$ is a multiple of the K\"ahler form,
it is of type-(1,1) and  the above condition is satisfied.
This is only  a local necessary condition; the existence of globally
holomorphic sections
is a  harder question. We will in fact construct holomorphic sections
in our examples.
This way we can reduce the
size of our Hilbert space and restrict ourselves to
holomorphic sections. We define the quantum Hilbert space ${\cal H}_Q$ as
the space of holomorphic sections of the prequantum line bundle.
There is a projection
$K$ from ${\cal H}_{Pre}$ into ${\cal H}_Q$ and a given prequantum operator
$\tilde f$ can be projected to an operator $\hat f=K\tilde f K$
acting on ${\cal H}_Q$.
Although $\hat f$ operates on the correct Hilbert space, they no longer satisfy
equation (\ref{qua}) in general.
We will, in fact, use another prescription to get the quantum operator,
which differs from this one by higher order terms.

In the language of geometric quantization, we will
obtain the quantum operators corresponding to
the moment maps and also see that
the above  commutation relations correspond to
the unitary representations of the central extensions of the unitary and
pseudo-unitary Lie algebras.

We have seen that $D_1$ is a contractible complex manifold.
The prequantum line bundle has a connection on it, we choose
the same expression as in the finite dimensional case,  given by
\beq
    \Theta={1 \over \hbar}
(\Tr (1-Z^{\dagger}Z)^{-1}dZ^{\dagger}Z-\Tr (1-Z^{\dagger}Z)^{-1}Z^{\dagger}dZ)
\eeq
Note that all the traces and inverses are well-defined here.
It also satisfies  $d\Theta=-{i \over \hbar} \omega$;
in the above coordinate system $\omega=-2i\partial_Z\partial_{Z^\dagger}
\log {\rm det}(1-Z^{\dagger}Z)$; since $Z \in {\cal I}_2$ the expression inside
 the determinant is of type $1 + {\cal I}_1$ and hence is well-defined.
We  use the homogeneity of $\omega$ and the equality of the two expressions
at $\Phi=\epsilon$, or $Z=0$ to fix the normalization.

The quantum Hilbert space is given, as before,
 by the holomorphicity requirement;
\beq
    {\cal H}_Q=\{ \psi | \nabla_{Z^{\dagger}} \psi=0,  \quad \psi \in
    {\cal H}_{Pre} \}
.\eeq
which has
solutions as $\psi={\rm det}^{1 \over \hbar}(1-Z^{\dagger}Z)\Psi(Z)$
, where $\Psi(Z)$ is an ordinary holomorphic function of the variables $Z$.

It is possible to establish an inner product on this space (and by completion,
turn it into a Hilbert space) following  ideas of G. Segal \cite{segal}.
 Alternatively, we can establish an inner product by Pickrell's measure 
 on the Grassmannian \cite{pickrell}
 which should also have a counterpart on the Disc.
 We will not address this issue in this paper. Of course in the finite 
dimensional 
 cases, standard measures exist which can be used to construct the Hilbert 
spaces.

We will obtain the action of the prequantum operators corresponding to
the moment maps on  $\cal H$.
Suppose that $f_{-u}$, where $u \in {\underline U}_1({\cal H}_-,{\cal H}_+)$,
 is the moment map and $V_u=V^Z_u\partial_Z+V^{Z^{\dagger}}
_u\partial_{Z^{\dagger}}$ is the vector field generated
by it.  We  calculate $\tilde f_{-u}$ action on the wave function
$\psi={\rm det}^{1 \over \hbar}(1-Z^{\dagger}Z)\Psi(Z)$.
We  note that the action of $V_u$ on $Z$ is given by;
${\cal L}_{V_u} Z=\alpha Z+\beta-Z\gamma Z-Z\delta$ where,
\beq
       u=\pmatrix{\alpha & \beta\cr
                  \gamma & \delta\cr}
\nn\eeq
is the decomposition of $u$ into block form; $\alpha^{\dagger}=-\alpha$
, $\beta^{\dagger}=\gamma$ and $\delta^{\dagger}=-\delta$ and further
$\gamma,\ \beta \in {\cal I}_2$. If we use the
moment map as $\Tr uM$
we can  write the action explicitly as;
\bea
    \tilde f_{-u}\psi=&-&\!i\hbar[-{2 \over \hbar}\Tr(1-Z^{\dagger}Z)^{-1}
     Z^{\dagger}(\alpha Z+\beta-Z\gamma Z-Z\delta)]\psi\nn\\
       &+&\! 2i[\Tr\alpha(1-(1-ZZ^{\dagger})^{-1})-\Tr\beta Z^{\dagger}
             (1-ZZ^{\dagger})^{-1}+\Tr \gamma(1-ZZ^{\dagger})^{-1}Z\nn\\
           &+&\!\Tr\delta Z^{\dagger}(1-ZZ^{\dagger})^{-1}Z)]\psi
       -i\hbar {\rm det}^{1 \over \hbar}(1-Z^{\dagger}Z)
      {\cal L}_{V_u}\Psi(Z)
.\eea
Note that, because of the vacuum subtraction, all the  traces in the above
expression are well-defined.
A careful calculation is needed to verify that at each step all the
traces are convergent. This can indeed be done with the result;
\beq
    \tilde f_{-u}\psi(Z,Z^{\dagger})=
   {\rm det}^{1 \over \hbar}(1-Z^{\dagger}Z)(-i\hbar)[{\cal L}_{V_u}\Psi(Z)
   -{2 \over \hbar}\Tr(\gamma Z) \Psi]
.\eeq
This differs from the finite dimensional case by a constant term.
We see that the trace in this formula is finite. Furthermore, the
changes introduced are all holomorphic.
This shows that the action of the moment maps in fact preserves the
holomorphicity requirement.
One can define a representation of the central extension of
 ${\underline U}_1({\cal H}_-,{\cal H}_+)$ on
the space of holomorphic functions
generated by $\tilde f_{-u}$ ;
\beq
   \tilde f_{-u}\Psi=-i\hbar[{\cal L}_{V_u}\Psi -{2 \over \hbar}
       \Tr (\gamma Z)\Psi]
.\eeq
This
can be exponentiated to a group action given by;
\beq
     \rho(g^{-1})\Psi={\rm det}^{-{2 \over \hbar}}(d^{-1}cZ+1)
                \Psi\big((aZ+b)(cZ+d)^{-1}\big)
.\eeq
In finite dimension this action gives a representation of the group;
but due to the vacuum subtraction it is not clear that the above
action gives a representation.
In fact if we calculate $\rho(g_2^{-1})\rho(g_1^{-1})\Psi(Z)$
and compare this with $\rho((g_1g_2)^{-1})\Psi(Z)$, after some
calculations, we see that they differ by a factor;
\beq
     c(g_1,g_2)={\rm det}^{2 \over \hbar} [(d_1d_2)^{-1}c_1b_2+1]
.\eeq
This  is well defined since $c_1b_2 \in {\cal I}_1$. We will see
that the representation we have
  corresponds to the representation of the central extension
of $U_1({\cal H}_-,{\cal H}_+)$.
Let us recall that a central extension of a group
$G$ by ${\bf C}^*$ is given by the following exact sequence;
\beq
      1 \to {\bf C}^* {\buildrel i \over \longrightarrow}
           {\hat G} {\buildrel \pi \over \longrightarrow} G \to 1
\eeq
 where  $i$ and $\pi $ are group homomorphisms. The extension can be
nontrivial  algebraically and also topologically, if the above sequence also
generates a nontrivial principal fiber bundle.
If the  extension is topologically trivial, as
in the case of $U_1({\cal H}_-,{\cal H}_+)$,
 there is an equivalent description.
In this case we  can find a globally defined map
$s : G \to {\hat G}$ such that $\pi(s(g))=g$.
Note that ${\hat G} \approx G \times {\bf C}^*$ as a topological
space,
and ${\bf C}^*$ is identified
with its image in the center of ${\hat G}$.
Thus we can think of $s$ as $s(g)=(g,\lambda(g))$; and see that
$s(g_1)s(g_2)=(g_1g_2, \lambda(g_1)\lambda(g_2))=(g_1g_2, c^{-1}(g_1,g_2)
\lambda(g_1g_2))$;
here $c$ is   a map $c:G \times G \to {\bf C}^*$ which measures
how much $s$ differs from a group homomorphism.
For this  $c$ must satisfy the so-called co-cycle condition;
$c(g_1g_2,g_3)c(g_1,g_2)=c(g_1,g_2g_3)c(g_2,g_3)$. An extension
will be algebraically non-trivial if there is no function $\phi:G\to {\bf C}^*$
, such that $c(g_1,g_2)=\phi(g_1)\phi(g_2)(\phi(g_1,g_2))^{-1}$.
We can see that our formula for $c$ satisfies the cocycle condition:
\bea
    c(g_1g_2,g_3)c(g_1,g_2)&=&c(g_1,g_2g_3)c(g_2,g_3)\nn\\
     &=&{\rm det}^{2 \over \hbar}[(d_1d_2d_3)^{-1}c_1a_2b_3
       +(d_2d_3)^{-1} c_2b_3+(d_1d_2)^{-1}c_1b_2+1].
\eea
note that the expression inside the square brackets is of type
$1+{\cal I}_1$, hence the determinant makes sense.
The cocycle $c$, in the finite dimensional case, can be
obtained from $\phi(g)=$det$(d)$.
This is not well-defined in  infinite dimensions; in fact,
the extension is nontrivial. Thus, we obtain a
representation of
a central extension $\hat{U}_1({\cal H}_-,{\cal H}_+)$ in
the Quantum Hilbert space of holomorphic sections.

Now we continue with the case of $Gr_1$, it turns out to
be  better to view  the Grassmannian as the coset
of yet another pair of groups.
This is because the extension is nontrivial both topologically
and algebraically. The cocycle does not exist as a continuous function.
 Essentially, we will enlarge  `numerator'
and `denominator' by the same amount so that the quotient is still the
Grassmannian.
In terms of these larger groups, we can find an explicit description of the
cocycle.
Let us
consider first the picture in terms of unitary  Lie groups.
Define

\beq
     \tE_1=\{(g,q)|g\in U_1({\cal H});q\in U({\cal H}_-); g_{11}q^{-1}-
1\in {\cal I}_1\},
\eeq
where we use the same decomposition as in (\ref{sp1}).
Group multiplication is just the pairwise product:
\beq
     (g,q)(g',q')=(gg',qq').
\eeq
Define also,
\beq
     \tilde F_1=\{(b,q)|b\in U({\cal H}_-)\times U({\cal H}_+); q\in
U({\cal H}_-);b_{11}q^{-1}-1\in {\cal I}_1\}
\eeq
so that the quotient remains the same:
\beq
     Gr_1=\tE_1/\tilde F_1.
\eeq
Now, we can construct a line bundle over $Gr_1$ starting
with a representation of $\tilde F_1$ on $\bf C$. We choose the
representation,
\beq
     \rho(b,q)={\rm det}^{N_c}[b_{11}q^{-1}].
\eeq
The determinant exists due to the condition  $b_{11}q^{-1}-1
\in {\cal I}_1$.
Now we see the reason for enlarging $GL_1$ and $B_1$.
 The determinant $\det b_{11}$ does not
exist in general; we need to factor out a part of it. This
is the role of the operator $q$.
The line bundle $\tilde L_\rho$ is  defined as
\beq
     \tilde L_\rho=(\tE_1\times_\rho {\bf C})/\tilde F.
\eeq
A section of this line bundle is then  a function
\beq
     \psi:\tE_1\to {\bf C}
\eeq
such that
\beq
     \psi(gb,qr)=\rho(b,r)\psi(g,q).
\eeq
An example would be the function
\beq
     \psi_0(g,q)={\rm det}^{N_c}[g_{11}q^{-1}].
\eeq
Of course if we  were to restrict to finite dimensions, 
the
dependence of $\psi$ on the  additional variable $q$ is just
an overall factor of $\det q^{-1}$ due to the equivariance
condition. Thus $\tilde L_\rho$ can be identified as the pre--quantum 
line bundle.

This new point of view on the $Gr_1$   forces us to re--examine the
classical theory. It is best to view $Gr_1$ as a coset space
of $\tE_1$ there as well. The action of $\tE_1$ on $Gr_1$
is just
\beq
     (g,q)\Phi\to g\Phi g^{-1};
\eeq
the new variable $q$ has no effect on $\Phi$. However, the
moment maps are now defined in terms of the Lie algebra of
$\tE_1$. Consider the function,
\beq
     f_{(u,r)}(\Phi)=2\Tr_\eps\bigg[\big({1-\Phi\over
2}\big)u\big({1-\Phi\over 2}\big)-\pmatrix{r&0\cr0&0}\bigg].
\eeq
In the finite dimensional case, this will be just the
previous moment map except for a piece independent of
$\Phi$. We can view this as a `vacuum subtraction' of the
moment map. If the operators are decomposed into $2\times 2$
blocks, we can check that the conditional trace exists.

We can also view this as a function on the group $\tE_1$
invariant under $\tilde F_1$:
\beq
     f_{(u,r)}(g,q)=2\Tr_\epsilon[\big({1-\eps\over
2}\big)g^{\dag}ug\big({1-\eps\over 2}\big)-q^{-1}rq].
\eeq
It takes a careful calculation,  to check that this is in fact
invariant under ${\tilde F}_1$. This moment map induces the
Hamiltonian vector field generating the infinitesimal action
of $\tE_1$:
\beq
     -df_{(u,r)}=\omega(V_{f_{(u,r)}},.),\quad
V_{f_{(u,r)}}=i[u,\Phi].
\eeq

Now we can proceed with  finding the pre--quantum operators.
We start by looking for a connection on the principal bundle
\beq
     {\tilde F}_1\to \tE_1\to Gr_1.
\eeq
In finite dimensions there is a connection on this principal bundle induced from 
an invariant metric.
In general it is not  possible to use an invariant metric on the
total space to find the connection.  The traces in the inner
product, which is  used in finite dimensions, will diverge.
 Therefore, we instead postulate an
expression for the connection one--form and check that it
indeed satisfies the necessary conditions \cite{kobayashi}.

We can regard a vector field  $Y=(X,T)$ on $\tE_1$ as an
ordered pair of operator--valued function on $\tE_1$ generating a left action,
\beq
     Y(g,q)=(X(g,q),T(g,q)).
\eeq
We will have, $X_{11}(g,q)-T(g,q)\in {\cal I}_1$.
Now, define the connection using the right action,
\beq
     \Omega(Y)(g,q)=({i\over 4}
                  [\eps,[\eps,g^{\dag}X(g,q)g]_+]_+,q^{-1}T(g,q)q).
\eeq
This is valued in the Lie algebra of ${\tilde F}_1$. The
first component agrees with the  expression for the
connection one--form in finite dimensions coming from  
the invariant inner product on the total space. The second
component is chosen in  such a way that $\Omega$ satisfies
\beq
     \Omega(V_{(u,r)})=(u,r)
\eeq
on vertical vector fields. This will allow us to define the
covariant derivatives acting on the sections of $\tilde L_\rho$.
We will obtain the prequantum operators, corresponding to the moment
maps we constructed,  acting on ${\cal H}_{Pre}$.
We define the covariant derivative of $Y(g,q)=(X,T)$,
\beq
    \nabla_{(X,T)}\psi(g,q)={\cal L}_{(X,T)}\psi(g,q)-
     \underline\rho(\Omega[(X,T)])\psi(g,q)
,\eeq
where $\underline\rho$ refers to the infinitesimal
 form of the representation $\rho$.
This expression reduces to the covariant derivative we  would obtain
in the finite dimensional case.

Let us write down the prequantum operator corresponding to a moment map;
\bea
    \hat f_{(u,r)}\psi&=&\! -i\hbar\nabla_{V(u,r)}\psi+f_{(u,r)}\psi
       =-i\hbar{\cal L}_{(u,r)}\psi\nn\\
&-&\! \hbar N_c\Tr_\epsilon[\big({1-\eps\over
2}\big)g^{\dag}ug\big({1-\eps\over 2}\big)-q^{-1}rq]\psi\nn\\
           &+& \! 2\Tr_\epsilon[\big({1-\eps\over
2}\big)g^{\dag}ug\big({1-\eps\over 2}\big)-q^{-1}rq]\psi,   \label{q1.b}
\eea
here the Lie derivative is defined through ${\cal L}_{(u,r)}\psi(g,q)=
\lim_{t \to 0}{\psi((1+itu)g,(1+itr)q)-\psi(g,q) \over t}$, and preserves the
 determinant condition since $u-r \in {\cal I}_1$. Since $\hbar N_c=2$
the last two terms cancel out and we end up with
\beq
   \hat f_{(u,r)}\psi(g,q)=-i\hbar{\cal L}_{(u,r)}\psi(g,q)
.\eeq
These operators provide a representation of the central extension of the
Lie algebra $\underline U_1({\cal H})$, as we will see.

We introduce the quantum Hilbert space ${\cal H}_Q$
using holomorphicity.
First, we will show that the Grassmannian is
a complex manifold. The complexification of $U_1({\cal H})$ is given by
\beq
   GL_1({\cal H})=\{ \gamma| \gamma \in GL({\cal H}), \quad [\epsilon, \gamma]
  \in {\cal I}_2 \}
.\eeq
We define the closed complex subgroup $B_1$ of
$ GL_1({\cal H})$, as the set of $\beta$'s, such that $\beta$  has  the
 decomposition into the block form;
$\beta=\pmatrix{\beta_{11} &\beta_{12}\cr 0 & \beta_{22}\cr}$, where
$\beta_{11} :{\cal H}_- \to {\cal H}_-$ and similarly for the others.
Using the same argument in finite dimensions, we see that the Grassmannian
is a complex homogeneous manifold given by
\beq
   Gr_1=GL_1({\cal H})/B_1
.\eeq
Even though we have a complex structure, because of the divergences,
 to define holomorphic
line bundles,  we need to extend
the complex  general linear group $GL_1({\cal H})$ to another group
$\tilde G_1$:
\beq
     \tilde G_1=\{ (\gamma, q)| q \in GL({\cal H}_-) ;\quad \gamma \in
GL_1({\cal H}) ,\quad \gamma_{11}q^{-1}-1 \in {\cal I}_1 \}
.\eeq
Here, similar to the previous cases, $\gamma_{11}$ denotes the mapping
$\gamma_{11} :{\cal H}_- \to {\cal H}_-$ in the block form of the matrix
$\gamma \in GL_1({\cal H})$.
$\tilde G_1$ is a complex Banach-Lie group under the multiplication
$(\gamma, q)(\gamma',q')=(\gamma\gamma',qq')$.
 We introduce $\tilde B_1$, a closed complex subgroup of $\tilde G_1$ as;
\beq
   \tilde B_1=\{ (\beta, t)| \beta \in B_1, t \in GL({\cal H}_-),
                \beta_{11}t^{-1}-1 \in {\cal I}_1 \}
\eeq
There is an action of $\tilde B_1$ on $\tilde G_1$, and this action is
holomorphic too.
We enlarged $GL_1({\cal H})$ and  $B_1$ with the same set of elements, thus the
quotient is still the same;
\beq
    \tilde B_1 \to \tilde G_1 \to Gr_1
.\eeq
Now, we can introduce the holomorphic line bundle corresponding to the
representation
$\rho(\beta, r)={\rm det}^{N_c}(\beta_{11}r^{-1})$. Since $\beta_{11}$
is invertible
it is enough to require $N_c$ to be an integer for the holomorphicity.
We can denote the associated
line bundle as $(\tilde G_1 \times_\rho {\bf C}) /\tilde
B_1$.  A section of this line bundle can be identified with equivariant
functions:
\beq
    \psi:\tilde G_1 \to {\bf C} \quad {\rm such \ that \  }
   \psi(\gamma\beta, qr)=\rho(\beta, r)\psi(\gamma, q).
\eeq
One such function is $\psi_0(\gamma,q)={\rm det}^{N_c}[\gamma_{11}q^{-1}]$.
We see that for this function to be globally holomorphic we need to
restrict ourselves to positive values of $N_c$.
We can write down a general expression for the holomorphic sections;
imagine that we decompose $\gamma$ as
\beq
    \gamma=\pmatrix{ \gamma_{11}& \gamma_{12}\cr \gamma_{21} & \gamma_{22}\cr}
.\eeq
Assume that we label the rows of $\gamma_{11}$ in some basis as
$(0,-1,-2,...,-k,...)$ and also the rows of $\gamma_{21}$ as
$(1,2,...,k,...)$. We define a matrix $\gamma_A$ which consists of the
rows of $\gamma_{11}$ and $\gamma_{21}$. $A$ denotes the rows we pick as
a sequence $(a_1,a_2,...,a_k,...)$ such that this sequence differs from
$(0,-1,-2,...,-k,...)$ only for finitely many  $a_i$'s.
If we think  of $\gamma_A$ as  $\gamma_A:{\cal H}_-
\to {\cal H}$ and extend $\gamma_{11}$ trivially to a map from ${\cal H}_-$
to ${\cal H}$, then  $\gamma_A-\gamma_{11}$ is a finite
rank operator.
Because
$\gamma_A=\gamma_{11}+R_A$, where $R_A$ is the finite rank piece,
$\gamma_Aq^{-1}-1$ is also in ${\cal I}_1$. This  implies that the determinant
${\rm det}(\gamma_Aq^{-1})$
is still well-defined. If we consider a set of such sequences
$A_1, A_2,...,A_p$, we can construct
 a general   solution :
\beq
    \psi(\gamma,q)={\rm det}^{w_1}(\gamma_{A_1}q^{-1})
      {\rm det}^{w_2}(\gamma_{A_2}q^{-1})...{\rm det}^{w_p}(\gamma_{A_p}q^{-1})
.\eeq
where, $w_1+ w_2+...+w_p=N_c$ and  they  all must be positive integers
for the holomorphicity. This, in turn, fixes the value of $\hbar$ to be
a {\it positive } number, ${2 \over N_c}$.
These sections are not all linearly independent, but one can find
a linearly independent family among them. As we remarked before, a suitable
 completion 
of the set of such holomorphic
sections constitute the quantum Hilbert space ${\cal H}_Q$.

If we look back at the formula (\ref{q1.b}), 
giving the action of moment maps on
sections, we see that the moment maps act as Lie derivatives. When we
restrict them to the holomorphic sections, they preserve the holomorphicity
condition; since,
the Lie derivative ${\cal L}_{(u,r)}$
generates the  infinitesimal action on the left
by constant operators $(u,r)$.
It is possible to exponentiate this to a group action. Consider the
left action of $(\lambda, s) \in \tilde G_1$ on the holomorphic sections:
${\rm r}(\lambda, s)\psi(\gamma,q)=\psi(\lambda^{-1}\gamma,s^{-1}q)$.
Since $\lambda$ and $s$ are constant matrices, the product still
satisfies the holomorphicity and due to the left action equivariance
is also preserved. We consider the following diagram;
\beq
       \matrix{ \ &\ & 1 &\ & 1 &\ &\ &\ &\ \cr
             \ &\ & \downarrow&\ &\downarrow&\ &\ &\ &\ \cr
             \ &\ &SL^1 &\rightarrow&\tilde{T}_1&\ &\ &\ &\ \cr
             \ &\ &\downarrow&\ &\downarrow&\ &\ &\ &\ \cr
    1&\rightarrow&GL^1&\rightarrow&\tilde E_1&\rightarrow&GL_1&\rightarrow&1\cr
    \ &\ &\downarrow&\ &\downarrow&\ &|&\ &\ \cr
    1&\rightarrow&{\bf
C}^*&\rightarrow&\widehat{GL}_1&\rightarrow&GL_1&\rightarrow&1\cr
    \ &\ &\downarrow&\ &\downarrow&\ &\ &\ &\ \cr
    \ &\ &1 &\ &1 &\ &\ &\ &\  \cr}
\eeq
In the above diagram, $GL^1$ is the subgroup of operators in $GL({\cal H}_-)$
for which the determinant exists. $SL^1$ is its subgroup of operators with
determinant one. $\tilde T_1$ denotes the subgroup of
$\tE_1$, which consists of elements  of the form  $(1,q)$, with determinant
of $q$ is one. The first horizontal map is the identification of the
two groups. The second one is an exact sequence of groups; which comes
from the fact that the second map
is a natural  imbedding and, third is a projection to the first factor.
The first vertical sequence is exact; if we define the second map as the
imbedding and third to be the determinant. The second vertical
sequence is also exact if we define $\widehat{GL}_1=\tE_1/\tilde T_1$.
In the last horizontal sequence, we introduce the second
map in a way that makes the diagram commutative.
If we combine all this information, it follows that the second horizontal
sequence of maps is also exact; that is to say
 we have a  central extension
$\widehat {GL}_1$
of the group $GL_1$. If we look at the action of $\tE_1$ on the sections
we note that if we multiply  an element $(\lambda, s)$ with $(1, t)$ with
det$t=1$; the representation r$(\lambda,s)$=r$(\lambda,st)$.
This shows that the  representation-r
on the space of holomorphic sections, factors through $\tilde T_1$.
Hence, we have a representation of $\widehat{GL}_1$ on ${\cal H}_Q$.
If we restrict
ourselves to the real compact form we get a representation of the
central extension $\hat U_1({\cal H})$. This is the representation
generated by the moment maps.

We can check that, for an arbitrary function $F(\Phi)$ on the Grassmannian
the associated prequantum operator will not, in general,
 preserve the holomorphicity.
In the infinite dimensional case, projecting to the holomorphic subspace
by an operator $K$, is even  more complicated.
We are interested in polynomial functions of $\Phi$, or $M$;
in this case we can
 introduce a simpler  quantization scheme. We replace each
$M$ in the polynomial, by the quantum operator we get using the moment
maps. This replacement requires an operator ordering rule,
such as normal ordering,  since  the operators associated to
different matrix elements of $M$, typically, do not commute.
For quadratic functions of $M$ we can state the normal ordering rule in  
Fourier expansion:
\beq
     \stackrel{\textstyle \circ}{\circ}\hat M(p,q) \hat M(r,s)
     \stackrel{\textstyle\circ}{\circ}=\cases{ 
     \hat M(r,s)\hat M(p,q) & if 
     $p<q$ and $r>s$\cr
        \hat M(p,q)\hat M(r,s) & otherwise\cr}
 \eeq
Similarly for the variables $\hat {\tilde M}(p,q)$. This will ensure that the 
interaction will not change the enrgy by an infinite amount with respect to the 
 free Hamiltonian.
 
The quantum operators we obtain, for the polynomial functions, will
continue to preserve the holomorphicity condition. Hence we have a
prescription for quantizing polynomial functions of $\Phi$ through
operators acting on the space of holomorphic sections. The price
one pays is that, the commutators of operators no longer form a representation
of the Poisson algebra. The deviations are higher order in $\hbar$, and this
situation is common in quantum theories.

\section{Application to Scalar QCD}

In this section we will  apply some of the previous ideas to
scalar quantum chromodynamics. We will show that the large-$N_c$ limit
of this theory is a somewhat  unusual  classical theory, and its
   phase space  is
the infinite dimensional Disc, $D_1$.
We will obtain the equations of motion, and show that their linearization
will give the analog of the `t Hooft equation in QCD, first obtained in the
reference \cite{Tomaras}. The infinite dimensional Grassmannian, $Gr_1$, was 
shown
to be the phase space of fermionic QCD in \cite{2dqhd}, by the first author.
In that case also, the linearization of the equations of motion was shown to
give the `t Hooft equation. The equations of motion are nonlinear in general;
in the fermionic case, these equations have soliton solutions, and
they are interpreted as baryons of that theory. The estimate of their
masses is given through a numerical solution 
in \cite{2dbaryon} and by a variational argument in \cite{2dqhd}.

We start with the Hilbert space of square integrable functions
${\cal H}=L^2({\bf R})$. The elements of this space will be thought of as
functions of a real variable which has the physical interpretation of the null
component of momentum. (See below.)
There is a decomposition of ${\cal H}$  into ${\cal H}_-$ and ${\cal H}_+$
where ${\cal H}_-$ corresponds to the positive momentum components
and ${\cal H}_+$ the negative momentum components in
the Fourier  expansion.
They can be thought of as eigensubspaces of the operator
$\epsilon(p,q)=-{\rm sgn}(p)\delta(p,q)$.
(Notice that we have a  different sign convention here, as compared to the
previous section.)
We introduce  $D_1$, the set of operators  $Z:{\cal H}_+ \to {\cal H}_-$,
to be:
\beq
    D_1=\{ Z | \ \Tr Z^{\dagger}Z < \infty \quad 1-Z^\dagger Z>0 \}
.\eeq

{}From the previous discussion we know that it is  convenient to
introduce the
variables $M=\int \tilde M(p,q)\lambda(q,p)dpdq$.
Here, $M$ was defined through a non-linear transformation:
\bea
         M=\pmatrix{2 & 0\cr
                    0& 0\cr} -2\pmatrix{ (1-ZZ^{\dagger})^{-1}
                                 &-(1-ZZ^{\dagger})^{-1}Z\cr
                         Z^{\dagger} (1-ZZ^{\dagger})^{-1}
                      &-Z^{\dagger}(1-ZZ^{\dagger})^{-1}Z\cr}
\eea
where, the expression
 is  written in the decomposition ${\cal H}={\cal H}_- \oplus {\cal
H}_+$. We know that the Disc is a homogeneous symplectic complex
manifold under the action of $U_1({\cal H}_-,{\cal H}_+)$.
We can think of it as the phase space of a dynamical system. In that case
the fundamental Poisson brackets satisfied by the coordinates can be expressed
through:
\bea
   &\ &\{\tilde M(p,q),\tilde M(r,s)\}=\nn\\
     &\ &i(\tilde M(r,q)\epsilon(p,s)-
       \tilde M(p,s)\epsilon(r,q))
     +i(\epsilon(p,s)\delta(r-q)-\epsilon(r,q)\delta(p-s)).  \label{q1.a}
\eea
The above realization, using complex valued functions, has  a physical
interpretation in terms of free scalar field theory. If we use the
light cone coordinates, $x^\pm={x^0 \pm x^1 \over \sqrt 2}$, the
metric of two dimensional Minkowski space
 becomes $ds^2=2dx^-dx^+$. The Lorentz transformations in this
language can be written as $x^- \to e^\theta x^-$ and $x^+ \to e^{-\theta}
x^+$, where $\theta$ is the rapidity.
Here, we consider $x^-$ to be the `space' coordinate and $x^+$ to be the
`time'. That is, initial data are given on a surface $x^+=$constant; and
the equations of motion predict the evolution of the fields off this surface.
$L^2({\bf R};{\bf C})$ is the space of complex valued functions on
$x^-$, and the momentum variable refers to $p_-$.
Let us write down the action of the complex scalar field with $N_c$
components in this language:
\beq
    S=\int ( \partial_-\phi^*_\alpha\partial_+\phi^\alpha+
            \partial_+\phi^*_\alpha\partial_-\phi^\alpha-
        m^2 \phi^*_\alpha\phi^\alpha)dx^-dx^+
\eeq
We see that in the light cone formalism, the action is first order in the
time variable. The phase space
of the theory is the set of functions $\phi : {\bf R} \to {\bf C}^{N_c}$.
Poisson brackets satisfied by the variables $\phi$,
can be read off, if we think of $\partial_-$ as  the symplectic
form. Imposing antisymmetry, we can calculate the inverse of this operator,
and  obtain the relation:
\beq
   \{ \phi^\alpha(x^-,x^+), \phi^*_\beta(y^-,x^+) \}={1 \over 2}{\rm
sgn}(x^--y^-)
      \delta^\alpha_\beta
.\eeq
and the other Poisson brackets, such as,
$\{\phi,\phi\},\{\phi^\dagger,\phi^\dagger\}$ vanish.
The Hamiltonian, which describes the null component ${\rm P}_+$ of the
total momentum of the field is:
\beq
     H_0={m^2 \over 2}\int \phi^*_\alpha\phi^\alpha dx^-
.\eeq
The quantization of this system would replace functions by operators and
the Poisson brackets by commutators:
\beq
     [\hat \phi^\alpha(x^-),\hat \phi^\dagger_\beta(y^-)]=
     -i{1 \over 2} {\rm sgn}(x^--y^-)\delta^\alpha_\beta
.\eeq
and the other combinations vanish.
Note that the above commutator is invariant under conjugate-transpose.
It is convenient  to  expand $\hat \phi^\alpha$ into the Fourier modes;
we will denote $p_-$ as $p$
for brevity:
$\hat \phi^\alpha=\int_{-\infty}^{\infty} {a^\alpha(p) \over \sqrt{2|p|}}
e^{ipx^-}[dp]$
.  The hermitian conjugate of this expansion will give the
expansion for $\hat \phi^\dagger_\alpha$.

We note that in order to satisfy the commutation relations we need to
take
\beq
    [a^\alpha(p) , a^\dagger_\beta(q)]=
            {\rm sgn}(p)\delta(p-q)\delta^\alpha_\beta
\eeq
and the other commutators vanish. We can build a bosonic Fock space
using the above set of operators. We define the
`vacuum' state $|0>$ as,
\beq
      a^\dagger_\alpha(p)|0>=0 \quad {\rm if}\ p \le 0
     \quad {\rm and} \quad a^\alpha(p)|0>=0 \quad {\rm if} \ p>0
.\eeq
 It is known that due to divergences
we need to use a normal ordering, and we define it as follows:
\beq
      :a^\dagger_\alpha(p)a^\beta(q):=\cases{ a^\dagger_\alpha(p)a^\beta(q)
                              & if $p>0$\cr
                 a^\beta(q)a^\dagger_\alpha(p) & if $p \le 0$ \cr}
\eeq
We can express the quantum Hamiltonian acting on this Fock space as:
\beq
      \hat H_0={m^2 \over 2}\int \sum_\alpha :a^\dagger_\alpha(p)
                  a^\alpha(p): {dp \over |p|}
\eeq
The equation of motion that follow from this is,
\beq
    (2\partial_+\partial_- +m^2)\hat \phi^\alpha=0
.\eeq
Let us introduce the variables $\hat{\tilde M}(p,q)={2 \over N_c}\sum_\alpha
:~a^\dagger_\alpha(p)a^\alpha(q)~:$,  satisfying the commutation relations
\beq
      [\hat {\tilde M}(p,q),\hat {\tilde M}(r,s)]={2 \over N_c}
   (\hat{\tilde M}(r,q)\epsilon(p,s)-
       \hat{\tilde M}(p,s)\epsilon(r,q)
     +\epsilon(p,s)\delta(r-q)-\epsilon(r,q)\delta(p-s)).
\eeq
These form a quantization of equation (\ref{q1.a}) if $\hbar={2 \over N_c}$.
One can also check by direct computation that this realization satisfies 
the quadratic 
constraint up to terms of order $1 \over N_c$ in the color invariant sector.
This is just a special case of what we discussed in section three. We can 
write down the equations of motion for the free field in terms of the 
variable $\hat{\tilde M}(p,q)$:
\beq
    \partial_+ \hat{\tilde M}(k,l;x^+)=i {m^2 \over 2}
    [{1 \over l}-{1 \over k}]\hat {\tilde M}(k,l;x^+)
.\eeq
In the classical limit,  $N_c \to \infty$,
$\hat{\tilde M}(p,q)$ tends to the classical variable $\tilde M(p,q)$
satisfying,
\beq
   {\partial \tilde M(k,l;x^+) \over \partial x^+}={i \over 2}m^2
    [{1 \over l}-{1 \over k}]\tilde M(k,l;x^+)
.\eeq
This is the same as the equation of motion we get through the Poisson
brackets if we take the Hamiltonian $H_0$, a   linear function
of  $\tilde M(p,q)$, as:
\beq
     H_0={m^2 \over 2}\int \tilde M(p,q) {\delta(p-q) \over \sqrt{|p||q|}}
                      dpdq   \label{q1.6}
.\eeq

One can, in fact, motivate this choice of the Hamiltonian independently.
We note that since the Poisson brackets are at equal null time $x^+$,
they must be Lorentz invariant. This implies that
 $\tilde M(p,q) \to e^{-\theta} \tilde M(e^{-\theta}p,e^{-\theta}q)$
under Lorentz transformations.
Since the  Hamiltonian  is the null component of the total
momentum of the field, it
must  transform as  $H_0 \to e^\theta H_0$.
Let us write $H_0=\int h(p,q)\tilde M(p,q)dpdq$, where we choose 
$h(p,q)=h(p)\delta(p-q)$ when we require  translation invariance (or momentum
conservation).
This implies that  $h(e^{-\theta} p)=e^\theta h(p)$ under Lorentz 
transformations.
Since the original Poisson brackets are invariant under complex conjugation, we 
have 
$\tilde M(p,q)^*=\tilde M(q,p)$. If we impose the reality of the Hamiltonian we 
should have 
$h(p)$ real.
If we also assume the parity invariance this implies that $\tilde M(p,q)$ 
must transform under parity as $\tilde M(p,q) \to \tilde M(-q,-p)$.
Thus, the parity invariance gives us
$h(p)=h(-p)$, and combining all these we see that $h(p)\sim {1 \over |p|}$.
This implies that $H_0={m^2 \over 2}\int \tilde M(p,q)
{\delta(p-q) \over \sqrt{|p||q|}} dpdq$ where we put a constant of
proportionality $m^2 \over 2$. This is a moment map, as we have seen
in the second section.

We see that a linear function of $\tilde M(p,q)$ (a moment map)
corresponds to free field theory. The same equation of
motion in the language of $Z$
is  highly nonlinear. Thus, even the free field theory has a complicated
description in terms of $Z$.

{}From our discussion we conclude that the two quantum theories,
quantization on $D_1$ and complex  scalar free field with
$N_c$ components in the color invariant sector,  are identical.
The large-$N_c$ limit, being a classical theory, is rigorously
defined through the classical dynamics on the infinite dimensional Disc, $D_1$.

The next question in this direction is to introduce interactions, and see
its meaning in the more conventional point of view.
The general interaction term one can think of
can be written as an integral kernel;
\beq
      H_I=\int \tilde G(p,q;s,t)\tilde M(p,q)\tilde M(s,t) dpdqdsdt
.\nn\eeq
There is an  immediate symmetry, $\tilde G(p,q;s,t)=\tilde G(s,t;p,q)$.
Translational invariance will require $\tilde
G(p,q;s,t)=G(p,q;s,t)\delta(p+s-q-t)$.
Lorentz invariance implies that
$G(p,q;s,t)=e^{-2\theta}G(e^{-\theta}p,e^{-\theta}q;e^{-\theta}s,e^{-\theta}t)$.
Using the transformation of $\tilde M(p,q)$, reality condition implies that
 $G(p,q;s,t)=G^*(q,p;t,s)$.
It is also reasonable to demand it to be non-separable, that is to say we cannot 
write the interaction as a sum of terms which are not related to each others by 
any symmetry.
Parity invariance puts a further restriction: $G(p,q;s,t)=G(-q,-p;-t,-s)$.
We can see that these are satisfied for  a simple non-separable kernel;
\beq
   H_I=\tilde \lambda\int {1 \over \sqrt{|p||q||s||t|}}\delta(p+s-q-t)\tilde 
M(p,q)
        \tilde  M(s,t) dpdqdsdt
        \label{be1}
.\nn\eeq
The choice of $\delta$-function is due to the momentum conservation; and the
momentum dependence is  the one necessary  for the correct Lorentz
transformation property.
This, in fact, corresponds to the large-$N_c$ limit of
$\lambda \phi^*_\alpha\phi^\alpha\phi^*_\beta\phi^\beta$
theory, with appropriate rescalings of the
coupling constant $\tilde \lambda$. 
Let us see this in more detail.

We assume that the classical theory is given by the free Hamiltonian and the 
following interaction added to it:
\beq 
     H_I=\lambda \int \phi^*_\alpha\phi^\alpha\phi^*_\beta
     \phi^\beta dx^-
\eeq
To write down the quantum theory we need to use the normal ordering, and it 
is better to express it in 
terms of the creation and annihilation operators:
\beq
    \hat H_I={\lambda \over 4}\int_{-\infty}^{\infty}  { 1 \over 
\sqrt{|p||q||s||t|}}
    \delta(p+s-t-q)\nn\\
      :~a^\dagger_\alpha(p)a^\alpha(q)a^\dagger_\beta(s)a^\beta(t)~:dpdqdsdt,
       \label{e1}
\eeq
where we use the same conventions as in the free field theory.
We normalize the free Hamiltonian and the interaction by $1 \over N_c$, and 
redefine the coupling constant ${\lambda \over 8}N_c=\tilde\lambda$, and 
take the limit 
$N_c \to \infty$ while keeping $\tilde \lambda$ fixed. In this limit 
the normal ordered 
product decomposes as a product of two normal ordered color invariant  
operators.
More explicitly, as $N_c \to \infty$,
 the normal ordered products
${1 \over N_c^2}:~a_\alpha^\dagger(p)a^\alpha(q)a_\beta^\dagger(s)a^\beta(t)~:$
decompose into ${1 \over N_c}:~a^\dagger_\alpha(p)a^\alpha(q)~:{1 \over N_c}
:~a^\dagger_\beta(s)a^\beta(t)~:+O({1 \over N_c})$. If we introduce the variable 
$\hat{\tilde M}(p,q)={2 \over N_c} :~a^\dagger_\alpha(p)a^\alpha(q):~$, in 
the limit 
$N_c \to \infty$ the interaction Hamiltonian (\ref{e1}) goes to 
the expression in (\ref{be1}). The constarint in the variable 
$\tilde M$ implies that we are looking at the color invariant sector of this 
theory. This is unlike the case of 2d QCD where the Hamiltonian diverges
except in the color invariant sector of the theory.
One can calculate the equations of motion for the basic variables $\tilde 
M(k,l)$, 
for this choice of the Hamiltonian:
\bea
 \partial_+\tilde M(k,l)&=&\!\!\!{m^2 \over 2}[{1 \over l}-{1 \over k}]\tilde 
M(k,l)\nn\\
  &+&\!\!\!2i\tilde \lambda \int \big[ {{\rm sgn}(k) \over 
\sqrt{|k(k+p)|}}\tilde 
  M(k+p,l)-{{\rm sgn}(l) \over \sqrt{|l(l+p)|}}\tilde M(k, l+p)\big]\times\nn\\
   &\ &\ \ \times\tilde M(s-{p \over 2},s+{p \over 2})
   {dpds \over \sqrt{|(s+{p \over 2})(s-{p \over 2})|}}\nn\\
  \!\! &+&\!\!\!2i\tilde \lambda {[{\rm sgn}(k)-{\rm sgn}(l)]\over 
\sqrt{|kl|}}\int 
    \!\tilde M(p+{k-l \over 2},p-{k-l \over 2})
     {dp \over \sqrt{|(p +{k-l \over 2})(p-{k-l \over 2})|}}.\nn\\
\eea
Later on we will comment on the linear approximation to this model.
  
Another choice, consistent with the Lorentz transformation property,
translational invariance and the parity invariance is,
\bea
    H_I={\tilde g^2 \over 2}\int_{-\infty}^{\infty}&\!\ \!\!\!&\delta(q+t-p-s)
     ({1 \over (p-t)^2}+{1 \over (q-s)^2}){qt+ps+st+pq\over
\sqrt{|p||q||s||t|}}\nn\\
      &\ &\tilde M(p,q)\tilde M(s,t)dqdpdsdt. \label{q1.5}
\eea
We can calculate the equations of motion by using the Poisson brackets
when  we add the
free part to this interaction Hamiltonian.
The result can be written as,
\bea
  \partial_+\tilde M(k,l)&=&\!i{m^2 \over 2}[{1 \over l}-{1 \over k}]\tilde
      M(k,l)\nn\\
   &+&\!i\tilde g^2\int\big [ \tilde M(p,l){\rm sgn}(k)\delta(p+s-k-t)
  { kt+ps+st+pk \over \sqrt{|pkst|}}({1 \over (p-t)^2}+{1 \over (s-k)^2})\nn\\
 &-&\!\tilde M(k,p){\rm sgn}(l)\delta(p+t-l-s){pt+ls+st+lp \over \sqrt{|lpst|}}
  ({ 1 \over (p-s)^2}+{1 \over (l-t)^2}) \big ]\tilde M(s,t)dpdtds\nn\\
  &+&\!i \tilde g^2\int ({\rm sgn}(l)-{\rm sgn}(k)) \delta(l+s-k-t)
   { kt +ls+kl+st \over \sqrt{|lkst|}}\nn\\
  &\times&\!  ({1 \over (l-t)^2}+{1 \over (k-s)^2})
   \tilde M(s,t)dsdt
.\eea
One can simplify the equations of motion by some redefinition of the variables:
\bea
  \partial_+\tilde M(k,l)&=&\!\!\!i{m^2 \over 2}[{1 \over l}-{1 \over k}]\tilde
      M(k,l)\nn\\
 + 2i\tilde g^2\int\!\!\!\!&\big[ &\!\!\!\!\tilde M(k+v,l){{\rm sgn}(k) \over 
   \sqrt{|k(k+v)|}}{(s-v/2+k)^2 \over (s-v/2-k)^2}-\tilde M(k,l-v){{\rm sgn}(l) 
\over 
    \sqrt{|l(l-v)|}}{(s+v/2+l)^2 \over (s+v/2-l)^2}\big]\times\nn\\
   &\ &\  \ \times {\tilde M(s-v/2,s+v/2) \over \sqrt{ 
|(s-v/2)(s+v/2)|}}dsdv\nn\\
    &+&\!\!\!2i \tilde g^2\int{{\rm sgn}(k)-{\rm sgn}(l) \over \sqrt{|kl|}}
     {[s+(k+l)/2]^2 \over [s-(k+l)/2]^2} {\tilde M(s-(l-k)/2,s+(l-k)/2)
     \over \sqrt{|(s-(l-k)/2)(s+(l-k)/2)|}}ds.\nn\\
\eea  
This is a rather complicated, nonlinear equation.
Using the above  form we can show that in fact Tr$M$ is conserved by the 
time evolution,  so we can think of it as a conserved charge which has
 the meaning of baryon number.
We recall that in two dimensional QCD Tr$M$ is a topological invariant and hence 
conserved by the equations of motion.

One can see that the above theory in more conventional language
corresponds to the large-$N_c$ limit
of a complex scalar field with a current-current interaction through
 a propagator.
We can write  the quantum theory, using the same conventions as in the
case of free field;
\beq
   \hat H ={1 \over 2}
    m^2\int_{-\infty}^{\infty}:a_\alpha^\dagger(p)a^\alpha(p):
    {dp \over |p|} + {1 \over 4}g^2\int_{-\infty}^{\infty}:\hat 
j^\alpha_\beta(x^-)
                |x^--y^-|\hat j^\beta_\alpha(y^-):dx^-dy^-  \label{q1.c}
.\eeq
here $\hat j^\alpha_\beta(x^-)=i:\partial_-\hat\phi^*_\alpha(x^-)\hat 
\phi^\beta(x^-)-
\hat \phi^*_\alpha(x^-)\partial_-\hat \phi^\beta(x^-):$
denotes the current and we sum over repeated indices. In terms of the creation
and 
annihilation operators, the normal ordered product of two currents  
has the following term;
\beq
   :[pa^\dagger_\alpha(p)a^\beta(t)+ta^\dagger_\alpha(p)a^\beta(t)]
    [sa^\dagger_\beta(s)a^\alpha(q)+qa^\dagger_\beta(s)a^\alpha(q)] :
    \label{e2}
\eeq    
Dividing $\hat H$ by $N_c$ and   redefining  the coupling constant
 as $g^2N_c \to g^2$, we can take the large $N_c$
limit. In this limit, the normal ordered expression (\ref{e2})
 actually splits into a product of
four color singlet operators, up to corrections of order $1 \over N_c$, 
as in the previous case.
With the identification $\hat{\tilde M}(p,q)={2 \over N_c}:~a^\dagger_\alpha(p)
a^\alpha(q)~:$,
as $N_c \to \infty$, $\hat H$ tends  to the expression in
(\ref{q1.5}) plus the free part (\ref{q1.6}).

It is possible  to obtain this Hamiltonian from the scalar QCD, by eliminating
the gauge degrees of freedom.
Let us consider the spin zero bosonic matter fields, $\phi^\alpha$,
which are in the fundamental representation of $U(N_c)$.
We write down the Lagrangian in $1+1$ dimensions, for $\phi$ coupled to
$U(N_c)$ Yang-Mills theory. $A_\mu$ is the Yang-Mills field, a Lie algebra
valued vector. It can be written as $A_\mu=A_\mu^a T_a$ where
$T_a$'s are the generators and $\alpha=1,...$, rank$U(N_c)$.
We use hermitian generators in the fundamental
representation, and normalize them with respect to the Killing form
$\Tr T_aT_b=\delta_{ab}$.
The action for Yang-Mills
fields, is defined through the field strength tensor $F_{\mu\nu}=\partial_\mu
A_\nu-\partial_\nu A_\mu+ig[A_\mu,A_\nu]$.
Here, commutator refers to the Lie algebra commutator. We also need to write
down the gauge invariant coupling, through the covariant derivatives;
$D_\mu\phi^\alpha=\partial_\mu\phi^\alpha+ig(A_\mu)^\alpha_\beta\phi^\beta$.
Similarly for the complex conjugate field, $\phi^*_\alpha$.
The action is given by,
\beq
    S=\int  (-{1 \over 4}\Tr F_{\mu\nu}F^{\mu\nu}+(D_\mu \phi)^{\dagger}
       (D^\mu\phi)-m^2\phi^{\dagger}\phi)dx^-dx^+
.\eeq
where, $\dagger$ denotes the conjugate-transpose of a matrix.
We write down the action in the light-cone coordinates,
and  we pick the light-cone gauge, $A^a_-=0$,
then the action becomes:
\beq
       S=\int dx^-dx^+({1 \over 2}(\partial_-A_+^a)^2+ig(\partial_-
         \phi^{\dagger}
       A_+\phi-\phi^\dagger A_+\partial_-\phi)-m^2\phi^\dagger\phi
+\partial_+\phi^\dagger\partial_-\phi+\partial_-\phi^\dagger\partial_+\phi)
.\eeq
We see that the $A_+$ has no time derivatives, thus it is not a dynamical
field. We can eliminate $A_+^a$ through its equation of motion.
We introduce $j^\alpha_\beta=i(\partial_-\phi^*_\alpha\phi^\beta-
\phi^*_\alpha\partial_-\phi^\beta)$;
and substitute the solution for $A_+^a$
then, we obtain the following action:
\beq
     S=\int (\partial_-\phi^*_\alpha\partial_+\phi^\alpha+\partial_+
      \phi^*_\alpha\partial_-\phi^\alpha)dx^-dx^+-\int(m^2\phi^\dagger\phi
       -{1 \over 2}g^2j^\alpha_\beta{1 \over \partial_-^2}j^\beta_\alpha
       )dx^-dx^+
.\eeq
We see that this action is first order in its `time' variable and it is
written completely in terms of the field $\phi$. We can read off the
Hamiltonian
\beq
     H=m^2\int dx^-\phi^*_\alpha\phi^\alpha +{1 \over 4}g^2
        \int j^\alpha_\beta(x^-)|x^--y^-|j^\beta_\alpha(y^-)dx^-dy^-
.\eeq
The quantization of this theory through conventional methods will lead
to the Hamiltonian (\ref{q1.c}). Therefore, the above choice of
the Hamiltonian has fundamental importance.

Let us remark that  the true large-$N_c$ theory, although it is
classical,  is not a conventional field theory. Its dynamical variable is
non-local and satisfies a complicated nonlinear equation.
In fact, the nonlinearity is what is necessary for the theory to accomodate 
soliton solutions. Baryons will arise as solitons of
 the above set of nonlinear equations. Mesons correspond
to  small oscillations around the vacuum configuration. As is shown by the
first author \cite{2dqhd}, this point of view is valid in the fermionic QCD in
$1+1$ dimensions.
The usual `t Hooft equation for the meson spectrum was obtained through
the linearization of equations of motion and by choosing a simple ansazt
 for the variable $M(p,q)$.
First, we will obtain the analog of `t Hooft equation, by
linearization.
Then we will show how to make a variational estimate for the  the baryon mass in 
scalar 
QCD.

To start linearization of the theory around
the vacuum configuration $\tilde M(p,q)=0$,
one should note that it is also necessary to linearize
the constraint equation, $M^2+[\epsilon, M]=0$.
Since we assume that $\tilde M(p,q)$ represents small oscillations, we
can disregard the quadratic term, explicitly:
\beq
     \int [dp][ds]\tilde M(p,s)
     [\lambda(s,p)\epsilon+\epsilon\lambda(s,p)](k,l)=0
.\eeq
This implies that,
\beq
    \tilde M(k,l)[1+{\rm sgn}(k){\rm sgn}(l)]=0
,\eeq
which is to say that,
$\tilde M(k,l)=0$ unless, $k>0$ and $l<0$, or $k<0$ and $l>0$.
 We start with the Hamiltonian $H=H_0 + H_I$ and calculate 
the Poisson bracket of $\tilde M(k,l)$ with $H$ in the linear
approximation.
To perform this,  we drop the quadratic terms in the equations
of motion, and write down the
interaction part only,
\bea
    \partial_{x^+}\tilde M(k,l)&=&\!{\tilde g^2}\int\delta(q-s+t-p)
     {1 \over (p-t)^2}{qt+ps+2st \over \sqrt{|p||q||s||t|}}
     ({\rm sgn}(l)-{\rm sgn}(k))\nn\\
    &\times&\! [\delta(p-l)\delta(q-k) \tilde M(s,t)+\delta(k-t)
     \delta(l-s) \tilde M(p,q)]dqdpdsdt
.\eea
We know that in the linear approximation $k$ and $l$ have the opposite
signs. We define $P=k-l$ and consider the case for which $k>0$ and $l<0$.
Let us concentrate on the first term, and perform the delta-function
integrations;
\bea
    I_1=\int {kt+l(P+t)+2t(P+t) \over \sqrt{(-kl)|t||P+t|}}
         {1 \over (l-t)^2}\tilde M(t+P,t)dt
.\nn\eea
Note that in the above expression, we can find the limits of integration,
since either $t+P>0$ and $t<0$, or $t+P<0$ and $t>0$. The second alternative
is impossible since $P>0$. We have $-P<t<0$ as the integration region.
If we make a change of variable  $t \to -t$ and redefine
$x={k \over P}$, $y={t \over P}$; we see that
\beq
   I_1={1 \over P}\int_0^1{(y-1)x+y(x-1)+2y(y-1) \over \sqrt{x(1-x)y(1-y)}}
                {1 \over (x-y)^2} \tilde M(Py,P(y-1))dy
.\eeq
The second term is very similar and we can apply the same set of
transformations to rewrite it as;
\beq
   I_2={1 \over P}\int_0^1{(y-1)x+y(x-1)+2x(x-1) \over \sqrt{x(1-x)y(1-y)}}
                {1 \over (x-y)^2} \tilde M(Py,P(y-1))dy
.\eeq
If we sum the two expression and insert to the resulting expression into
the full equation of motion, we obtain:
\bea
   \partial_{x^+}\tilde M(Px, P(x-1))&=&\!
   i{m^2 \over 2P}[{1 \over x} -{1 \over (x-1)}]\tilde M(Px,P(x-1))\nn\\
 &-&\! i{4 \tilde g^2 \over  P}\int^1_0{(x+y)(2-x-y) \over \sqrt{(1-x)x(1-y)y}}
     {1 \over (x-y)^2}\tilde M(Py,P(y-1))dy
\eea
The above form suggets that we define a quark-antiquark
wave function  $\xi(x)e^{iP_+x^+}=\tilde M(Px,P(x-1);x^+)$,
in a definite `energy' eigenstate $P_+$. We also express
the coupling constant in terms of coupling constant of  the gauge theory:
$\tilde g^2= {g^2 \over 32 \pi}$.
When this is substituted into the equations of motion, we see that:
\beq
   \mu^2 \xi(x)=
     m^2 [{1 \over x} +{1 \over 1-x}]\xi(x)
   -{g^2 \over 4\pi }\int^1_0{(x+y)(2-x-y) \over \sqrt{(1-x)x(1-y)y}}
     {1 \over (x-y)^2}\xi(y)dy
.\eeq
where, $\mu^2=2PP_+$ is the invariant mass of the meson; in the linear
approximation.
This is the equation obtained by Tomaras in \cite{Tomaras}, as the analog of
the `t Hooft equation for scalar QCD. The mass spectrum one will get from
the semiclassical techniques is similar to  the fermionic QCD.

We emphasize that  we have obtained  all the  nonlinearities in the large $N_c$
limit of scalar QCD. In principle we can obtain the interactions of the mesons
with each other by going to the next order in the linearization. This would be
the analogue of the work of Callan, Coote and Gross 
\cite{cal-gros} in scalar QCD. Our approach also describes  phenomena that 
cannot be 
 seen to any order of such an expansion, such as solitons. The simplest such 
soliton is 
 the baryon, more complicated ones describe the analogues of nuclei in scalar 
QCD.
 
 We can obtain an estimate of the baryon mass by a variational approach. If we 
 choose,
 ( as in \cite{2dqhd}) $M=-2uu^{\dag}$ where $u$ staisfies
  $\eps u=u,||u||^2=u^{\dag}u=1$, 
 we will satify all the constraints on $M$: $M^2+[\eps,M]=0,M^{\dag}=\eps 
M\eps$.
 Moreover, being rank one, it obviously satisfies the convergence conditions.
 In the variable $M$, the baryon number is just $-{1\over 2}\tr M$ which is 
equal to one 
 for our choice.

 In momentum space, $u$ will be represented by a function $\tilde u(p)$ which 
 vanishes for 
$p<0$.
 We can get a formula for the energy of this configuration by substituting it
  into our Hamiltonian. 
 The `kinetic energy' is best written in momentum space and the  `potential 
energy' in 
position space:
 \beq
 	H(v)=m^2\int_0^\infty  |\tilde v(p)|^2{[dp]\over p}+
 		\tilde g^2\int dx dy |x-y|\Im v^*(x)v'(x)\Im v^*(y)v'(y)
 		\eeq
 where $\Im$ refers to the imaginary part, 
  $\tilde v(p)={\tilde u(p)\over \sqrt p}$ and 
 $v(x)=\int_0^\infty e^{ipx} \tilde v(p)[dp]$. The minimum of  this over all 
$\tilde v$ 
satisfying $\int |\tilde v(p)|^2 p[dp]=1$ is the  mass of the lowest energy 
baryon, in the 
large $N$ limit. This can be found by numerically solving the resultant integral 
equation.
 A reasonable estimate can be made using a variational ansatz, such as $\tilde 
v(p)=C 
pe^{-p\kappa}$. ($C$ is a normalization constant.)

In the linear approximation the theory defined  by the interaction Hamiltonian
\cite{cole}
given in (\ref{be1}), reduces to a  similar integral equation,
\beq
    \mu^2 \xi(x)=
     m^2 [{1 \over x} +{1 \over 1-x}]\xi(x)
   +\tilde \lambda \int^1_0{1\over \sqrt{(1-x)x(1-y)y}}
    \xi(y)dy
.\eeq
with the same identifications for all the variables as before.
One can see that this equation can be solved 
analytically and it gives an equation for 
the square of the invariant mass of the small excitations, $\mu^2$:
\beq 
     {1 \over \mu\sqrt{4m^2-\mu^2}}{\rm arctan}{\mu \over \sqrt{4m^2-\mu^2}}=
     -{1 \over \tilde \lambda},
\eeq
where we assumed that the spectrum satisfies $\mu^2 < 4m^2$.
In addition to the discrete spectrum, this theory also has a continuous spectrum
of scattering states; unike in two-dimensional QCD, the particles are not 
confined.

\section{Acknowledgements}
We would like to thank S. Guruswamy and P. Vitale for  
discussions at the beginning stages, and to R. Henderson and C. W. H. Lee 
for reading the article and discussions. The second author would like to 
thank to  IHES,where he is presently an EPDI fellow, for the excellent
 environment 
provided during the completion of this work.
Our research is also supported by the grant DE-FG02-91ER40685. 

\section{Appendix}
In this appendix we will give a detailed 
presentation of the finite dimensional case, 
at the expense of repeating some of the previous discussion. 
We start with the example of a finite dimensional Grassmannian manifold.
\cite{chern}
To be self--contained we will collect together some basic facts about  the
Grassmannian.
For our purposes it is most convenient
to define the Grassmannian as a set of Hermitian matrices satisfying a
quadratic constraint;
\beq
     Gr_M(m)=\{ \Phi | \Phi^\dagger=\Phi, \Phi^2=1, \quad \Tr \Phi=M-2m\}
,\eeq
where $\Phi$ is an $M \times M$ matrix. It has eigenvalues $+1$ and $-1$.
Due to the trace condition,  $m$ of them are $-1$ and the remaining
$M-m$ are $+1$. Every hermitian matrix can be diagonalized by some
unitary matrix $g \in U(N)$, therefore $\Phi$ can be
written as  $g\epsilon g^\dagger$ where
\beq
     \epsilon=\pmatrix{ -1_{m \times m} & 0\cr
                         0 & 1_{M-m \times M-m}\cr}
.\eeq
The Grassmannian $Gr_M(m)$ is thus the orbit of $\eps$ under $U(M)$.
One should note that we will obtain the same matrix $\Phi$ by using
$gh$ instead of $g$ where $h$ is a unitary matrix which commutes with
$\epsilon$. The set of such submatrices is the subgroup  $U(m) \times U(M-m)$.
Therefore,  if we start from $\epsilon$ and act on it with $U(M)$,
the stability subgroup of $\epsilon$ is $U(m) \times U(M-m)$. This
defines the orbit of $\epsilon$ as a quotient of $U(N)$ with its closed
subgroup $U(m) \times U(M-m)$.
\beq
      Gr_M(m)={U(M) \over {U(m) \times U(M-m)}}
.\eeq
This also shows that $Gr_M(m)$ is a compact manifold. The action of
unitary group on $Gr_M(m)$ will be given by $\Phi \mapsto g\Phi g^{\dagger}$.
One can give a more geometric meaning to the $Gr_M(m)$; to each $\Phi \in
Gr_M(m)$ there is a subspace of ${\bf C}^M$ of dimension $m$; namely, the
eigensubspace of $\Phi$ with eigenvalue $-1$.  $Gr_M(m)$
can thus be viewed as the set of $m$ dimensional subspaces of ${\bf C}^M$.

This geometric
picture provides us with another description of $Gr_M(m)$ as a coset space.
The subspace corresponding to $\epsilon$ consists of vectors $ v \choose 0$,
where $v \in {\bf C}^m$. The stabilizer of this subspace under the action of
$GL(M,{\bf C})$ is the Borel subgroup;
\beq
     B_m=\left\{ \pmatrix{a &b\cr
                          0 &d\cr} \Bigm| a\in GL(m,{\bf C}) \quad
          d \in GL(M-m,{\bf C})\ {\rm and }\ b\in 
   {\rm Hom}({\bf C}^{M-m},{\bf C}^m)
             \right\}
.\eeq
Moreover, any $m$-dimensional subspace can be brought to this form
by an action of $GL(M,{\bf C})$. Thus we can think of the Grassmannian as a
coset space of $GL(M,{\bf C})$ as well;
\beq
      Gr_M(m)=GL(M,{\bf C}) / B_m
.\eeq
This  point of view shows  that $Gr_M(m)$ is  a complex manifold,
since it is the quotient of a complex Lie group by a closed complex subgroup.

It is possible to give an explicit coordinate system for the Grassmannian
in terms of $m \times (M-m)$ complex matrices $Z$, given by,
\beq
    \Phi(Z)=1_{M \times M}-2\pmatrix{
(1+ZZ^{\dagger})^{-1}&(1+ZZ^{\dagger})^{-1}Z\cr
Z^{\dagger}(1+ZZ^{\dagger})^{-1} & Z^{\dagger}(1+ZZ^{\dagger})^{-1}Z\cr}
\eeq
This is not a global coordinate system and we need  $M \choose  m$
different charts to
cover $Gr_M(m)$. Because of that the use of coordinate systems is not
efficient. In a given chart, $U(M)$ acts on $Z$ by fractional linear
transformations;
$Z \mapsto (aZ+b)(cZ+d)^{-1}$ where  $g \in U(M)$ is decomposed into
the block form
\beq
        g=\pmatrix{ a_{m \times m} & b_{m \times (M-m)}\cr
                  c_{(M-m) \times m} & d_{(M-m) \times (M-M)}\cr}
\eeq

We define the generalized Disc to be
the space of $m \times (M-m) $ complex matrices, $Z$,  which satisfy
the inequality $1_{(M-m) \times(M-m)}- Z^{\dagger}Z>0$.
This gives an open region in ${\bf C}^{m \times (M-m)}$, which is
contractible.
We see that $D_M(m)$ is a complex manifold  with a single
coordinate chart.

We define the Pseudounitary group as follows;
\beq
     U(m,M-m)=\{ g | \ g \in GL(M,{\bf C}) \quad g\epsilon
     g^{\dagger}=\epsilon \}
,\eeq
where $\epsilon$ is the matrix defined previously.
Any element $g$ of $U(m, M-m)$ has an
action on $D_M(m)$  given by,
\beq
     Z \mapsto (aZ+b)(cZ+d)^{-1}
\eeq
where
\beq
       g=\pmatrix{ a_{m \times m} & b_{m \times (M-m)}\cr
                  c_{(M-m) \times m} & d_{(M-m) \times (M-M)}\cr}
\eeq
is a decomposition of $g$ into block forms.
The action of $U(m,M-m)$ is transitive on the set of matrices with
 $1_{(M-m) \times (M-m)}-Z^{\dagger}Z>0$
and the
stability subgroup of $Z=0$ is given by $U(m) \times U(M-m)$.
This shows  that $D_M(m)$
is  a homogeneous space.
We have a description of $D_M(m)$ as
a quotient space;
\beq
     D_M(m)={U(m,M-m) \over U(m) \times U(M-m)}
\eeq

We  will provide a map from the Disc to a set of
pseudohermitian matrices $\Phi$; if we
define,
\beq
     \Phi=1_{M \times M}-2\pmatrix{ (1_{m \times m}-ZZ^{\dagger})^{-1}
                                 &-(1_{m \times m}-ZZ^{\dagger})^{-1}Z\cr
                         Z^{\dagger} (1_{m \times m}-ZZ^{\dagger})^{-1}
                      &-Z^{\dagger}(1_{m \times m}-ZZ^{\dagger})^{-1}Z\cr}
.\eeq
One can check that the above set of matrices satisfies the
properties,
\beq
      \Phi^{\dagger}=\epsilon \Phi \epsilon  \quad
      \Phi^2=1 \quad \Tr \Phi=M-2m
.\eeq
Under the action of $U(m,M-m)$ the $\Phi$'s transform as,
\beq
      \Phi \mapsto g^{-1}\Phi g \quad {\rm for} \quad Z \mapsto g \circ Z
      \quad  g \in U(m, M-m)
.\eeq
We will see that the classical dynamics is most natural
in this language. The other advantage is in  showing  the parallels with the
Grassmannian.

We will introduce classical dynamics on $Gr_M(m)$ and $D_M(m)$; there
is a symplectic form on each one given by,
\beq
       \omega= {i \over 4} \Tr \Phi d\Phi \wedge d\Phi
,\eeq
It is invariant under $U(M)$ for the Grassmannian and invariant under
$U(m,M-m)$ for the Disc as can be checked. Since the spaces are
homogeneous it is enough to check the non-degeneracy at the point $\epsilon$.
If we denote the
components of a tangent vector at point $\Phi$ as $V(\Phi)$,
it  has to satisfy
the equation $[V(\Phi), \Phi]_+=0$ which comes from the constraint $\Phi^2=1$.
We also need to have unitarity and pseudo-unitarity conditions respectively.
At $\epsilon$ this means that for the Grassmannian;
$V(\epsilon)=\pmatrix{0&v\cr v^{\dagger}&0\cr}$ and for the Disc;
$V(\epsilon)=\pmatrix{0&v\cr -v^{\dagger}&0\cr}$.
Contracting with $\omega$ at $\epsilon$ we get;
\beq
     \omega(V_1(\epsilon), V_2(\epsilon))={i \over 4}
        \Tr \epsilon [V_1(\epsilon), V_2(\epsilon)]= {i \over 2}
        \Tr (v_1v_2^{\dagger}-v_2v_1^{\dagger}).
\eeq
which is clearly non-degenerate. Incidentally, this demonstrates that
$\omega$ is of type (1,1) with respect to the complex structure.
Closedness of $\omega$  can be proved using;
\bea
      d\omega&=& { i \over 4}\Tr d\Phi d\Phi d\Phi=
        { i \over 4}\Tr d\Phi d\Phi d\Phi\Phi^2=
      - { i \over 4}\Tr \Phi d\Phi d\Phi d\Phi\Phi\nn\\
      &=&- { i \over 4}\Tr \Phi^2 d\Phi d\Phi d\Phi=-d\omega,\nn
\eea
where we used $\Phi d\Phi+d\Phi \Phi=0$ and the cyclicity of the trace.

Since both of these symplectic manifolds  are homogeneous, it is possible to
find
a generating function for the respective group actions.
The infinitesimal group action is given by,
\beq
     \delta \Phi= it[u, \Phi]
\eeq
where $t$ is an  infinitesimal parameter,
$u=u^{\dagger}$ for the unitary group and $u=\epsilon u^{\dagger}
\epsilon$ for the pseudounitary group.
The vector field generating the group action is $V_u(\Phi)=i[u, \Phi]$.
If we insert this to the equation $-df_u=i_{V_u}\omega$,
we get,
\beq
    - df_u= {i \over 4}\Tr \Phi [V_u(\Phi), d\Phi]
\eeq
and using ${1 \over 4}[\Phi,[\Phi,u]]=u$ one can see that
$df_u=-\Tr ud\Phi$. An immediate solution for this is given by
$f_u=- \Tr u\Phi$. These are called the moment maps.
We can calculate the Poisson brackets of the moment maps in both
cases and we see that;
\beq
   \{ f_u, f_v \}= f_{-i[u,v]}
.\eeq
They provide  a symplectic realization of the respective Lie algebras.
We can express them by using the explicit coordinates.
Define $\Phi=\Phi^i_j e^j_i$ for the Grassmannian and
$\Phi=\tilde \Phi^i_j \lambda^j_i$ for the Disc, where $e^i_j$'s are called
the Weyl matrices, they have matrix elements $(e^i_j)^k_l=\delta^i_l\delta^k_j$
and $\lambda^i_j=\epsilon^i_k e^k_j$.
Note that $\Phi^{* i}_{j}=\Phi^j_i$ and $\tilde \Phi^{* i}_{j}=\tilde
\Phi^j_i$ as reality conditions in these basis.
For the Grassmannian;
\beq
    \{\Phi^i_j,\Phi^k_l\}=-i(\Phi^i_l\delta^k_j-\Phi^k_j\delta^i_l)
     \label{q1.1}
.\eeq
and as for the Disc;
\beq
    \{ \tilde \Phi^i_j,\tilde \Phi^k_l\}=-i(\tilde \Phi^i_l
        \epsilon^k_j-\tilde \Phi^k_j\epsilon^i_l)
    \label{q1.2}
.\eeq
In our applications, the Hamiltonians of interest will be of the
form $E=\Tr (h \Phi +\hat G(\Phi)\Phi)$ where $h$ is hermitian and pseudo-
hermitian respectively. $\hat G(e^i_j)^k_l=G^{ik}_{jl}$ will
represent the interaction and
chosen such that $G^{ij}_{kl}=G^{* kl}_{ij}$
for the Grassmannian and for the Disc, $ G^{ij}_{kl}=
[ (\epsilon~\otimes~\epsilon) G (\epsilon~\otimes~\epsilon)]^{* kl}_{ij}$.

For the sake of completeness, we will give the solution to the equations of
motion when there is no interaction;
\beq
    { d\Phi^i_j \over dt}=i[h,\Phi]^i_j \qquad \rightarrow
                  \Phi(t)=e^{iht}\Phi(0)e^{-iht}.
\eeq
for the Grassmannian and as for the Disc,
\beq
    { d(\tilde \Phi\epsilon)^i_j \over dt}=i[h,(\tilde \Phi\epsilon)]^i_j
 \qquad \rightarrow \tilde \Phi(t)=e^{iht}\tilde \Phi(0)e^{-ih^{\dagger}t}.
\eeq
where we think of $\tilde \Phi$ as a hermitian matrix. Note that
$h$ is pseudohermitian and the exponential on the right is in fact the
conjugate of the one on the left. This preserves the hermiticity on $\tilde
\Phi^i_j$ and also the constraint.

At this point we would like to make a digression.
 The homogeneous symplectic manifolds $Gr_M(m)$ and 
$D_M(m)$ arise naturally in the theory of group representations and co-adjoint 
orbits. Let us define the coadjoint orbits: Let $G$ be a Lie group and 
$\underline{G}$ its Lie algebra. The vector space dual to
$\underline{G}$, is denoted by $\underline{G}^*$. There is an action of 
$G$ on $\underline{G}$ by conjugation, called the adjoint action:
$g \in G$ and $u \in \underline{G}$, then $u \to gu g^{-1}$.
We denote this by ${\rm ad}_g$.
One can define an action of $G$ on $\underline{G}^*$ by using the adjoint 
action:
\beq
      ({\rm ad}_g^*\xi,u) \equiv -(\xi, {\rm ad}_gu)
\eeq
where $\xi$ is in the dual space, and $(.,.)$ denotes the natural pairing.
Given any point in the dual space, there is an orbit corresponding to it under 
the co-adjoint action. The remarkable fact is that, these spaces have a 
symplectic form on them, and the resulting orbits are homogeneous symplectic 
manifolds of the group $G$.
The infinitesimal actions will lead to tangent vectors on the orbit; 
and we can think of the vectors  corresponding to the 
Lie algebra elements $u,v$ as ${\rm ad}^*_u$ and ${\rm ad}^*_v$
respectively. We define the symplectic form at the point $\xi$ to  be:
\beq
      \omega_\xi({\rm ad}^*_u,{\rm ad}^*_v)=-\xi([u,v]) \label{q1.3}
,\eeq
 which is well-defined since $\xi$ is in the dual.
One can check that this form is closed,  non-degenerate, and homogeneous
\cite{kirillov}.
For semisimple Lie algebras, one can identify the dual of the Lie algebra
with the Lie algebra itself as a vector space, using the Killing form. 
In this case co-adjoint orbits are the same as the orbits of Lie algebra
elements under the adjoint action. This, of course, is  not true in general.
We see  that the Grassmannian and the Disc  are 
both coadjoint orbits of the matrix $\epsilon$, under the unitary group 
$U(M)$ and the pseudounitary group $U(m,M-m)$ respectively.
The symplectic forms we defined agree with the symplectic form defined 
on  a coadjoint orbit by (\ref{q1.3})  up to a numerical factor.

Before we talk about the geometric quantization of the above systems we
would like to make a connection with the conventional algebraic
quantization. We will construct the respective fermionic and bosonic Hilbert
spaces and see that they carry a representation of the symmetry group.
A more detailed analysis of the Grassmannian   was given in \cite{2dqhd}.
For the Grassmannian define the fermionic creation and annihilation
operators satisfying;
\beq
    [\chi^\alpha_i, \chi^{\dagger  j}_\beta]_+=\delta^j_i\delta^\alpha_\beta
.\eeq
and all the other anticommutators vanish. Here, $i=1,...,M$ and $\alpha=
1,...,N_c$.  The index $\alpha$ labels a certain number of copies of otherwise
identical fermionic systems; it is often called a `color' index.
The fermionic Fock space $\cal F$, on which these act,
 is $2^{MN_c}$ dimensional.
We introduce the operators
$\hat \Phi^i_j={1 \over N_c}\sum_\alpha
[\chi_j^\alpha,\chi^{\dagger i}_\alpha ]$;
 one can check that they satisfy the commutation relations,
\beq
    [\hat \Phi^i_j, \hat \Phi^k_l]={2 \over N_c}
                (\delta^i_l\hat \Phi^k_j-\delta^k_j\hat \Phi^i_l).
\eeq
This can be viewed as a quantization of the Poisson Bracket relations
(\ref{q1.1})
with $2 \over N_c$ playing  the role of $\hbar$.
Note that $(\hat \Phi^i_k)^{\dagger}=\hat \Phi^k_i$, and this is a unitary
representation. This representation is reducible; the operator
$\chi_i^{\dagger\alpha}\chi_\alpha^i$ summed over both $\alpha$ and $i$
commutes with the $\hat \Phi^i_j$. Also, the color operators, which generate
an $SU(N_c)$ symmetry,
\beq
     Q^\alpha_\beta={1 \over 2}\sum_i([\chi^{\dagger i}_{\beta},
          \chi^\alpha_i]-{1 \over N_c}\delta^\alpha_\beta
         [\chi^{\dagger i}_{\gamma}, \chi^\gamma_i])
.\eeq
commutes with them. If,
therefore, we  limit ourselves to the fixed values of particle number and
zero color,
\beq
     {\cal F}_{0m}=\{ |\psi> | \quad Q^\alpha_\beta|\psi>=0
    \quad  {1 \over N_c}\chi_i^{\dagger\alpha}\chi_\alpha^i|\psi>=m|\psi>\}
.\eeq
we get an irreducible representation.
We can see that on this space  $\hat \Phi^i_i=M-2m$.
 One can also check that the quadratic
constraint is  satisfied up to terms of order $1 \over N_c$. This has to
be anticipated since $1 \over N_c$ has the role of $\hbar$ and quantization
of higher order operators can introduce ambiguities of order
$\hbar$ which disappears at the limit. Thus we conclude that the quantization
of the classical system on the Grassmannian with Poisson brackets (\ref{q1.2})
 leads to
the color singlet operators of a system of fermions with  $N_c$ colors. For
this interpretation to work, the quantum parameter $\hbar$ must be of the form
${2 \over N_c}$ for some integer $N_c$. Also the classical limit correspondes
to the
limit of a large number of colors.

We can give an analogous construction for the Disc, $D_M(m)$ based
 on the bosonic Hilbert space.
 Define bosonic creation and annihilation operators;
\beq
    [a^\alpha_i,a^{\dagger  j}_{\beta}]=-\delta^\alpha_\beta\epsilon^j_i
,\eeq
and all the other commutators vanish. Here $i,j=1,...M$ and
$\alpha,\beta=1,...,N_c$
We define the operators $\hat {\tilde \Phi^i_j}={1 \over N_c}\sum_\alpha
[a^{\dagger i}_{\alpha}, a^\alpha_j]_+$.
One can show that they satisfy the commutation relations,
\beq
    [ \hat{\tilde \Phi^i_j},\hat{\tilde \Phi^k_l}]={2 \over N_c}
      (\hat{ \tilde \Phi^k_j}\epsilon^i_l-\hat {\tilde \Phi^i_l} \epsilon^k_j)
.\eeq
In the same way, we see that $(\hat {\tilde \Phi^i_j})^{\dagger}=
\hat{ \tilde \Phi^j_i}$.
Thus, we have a {\it unitary} representation of the pseudo-unitary group.
One can see that the operator $a^\alpha_j a^{\dagger  k}_{ \alpha}
\epsilon^j_k$ commutes with all the $\hat {\tilde \Phi^i_j}$ and can be fixed
to get an irreducible representation. We have a color $SU(N_c)$
symmetry generated by,
\beq
     Q^\alpha_\beta={1 \over 2}\sum_i([a^{\dagger i}_{\beta},
          a^\alpha_j]_+-{1 \over N_c}\delta^\alpha_\beta
         [a^{\dagger i}_{\gamma}, a^\gamma_j]_+)\epsilon^j_i
.\eeq
$Q^\alpha_\beta$ commutes with all the $\hat{\tilde \Phi}$'s.
Thus we can define ${\cal B}_{0m}$
to be the space,
\beq
     {\cal B}_{0m}=\{ |w> | \quad Q^\alpha_\beta|w>=0
       \quad  {1 \over N_c}a^\alpha_j a^{\dagger k}_{\alpha}
           \epsilon^j_k|w>=-m|w> \}
\eeq
carrying an irreducible representation.
In this space the trace condition is satisfied  and
the quadratic constraint is  satisfied up to  terms of order $1 \over N_c$. Thus
quantization of the system whose phase space is the Disc $D_M(m)$ leads to the
color singlet sector of bosons.

These facts   can also be seen  from the point of geometric quantization.
We will proceed to finding the prequantum Hilbert space and reducing it
to
the Quantum Hilbert space by imposing  holomorphicity.
We will describe the case of  the  generalized Disc and be more concise in the
case of the Grassmannian.
The generalized Disc is a contractible space, therefore all line bundles on
it are topologically trivial and we can use a global coordinate system.
The sections of such a bundle are given by complex valued
functions on the Disc and they are all globally defined.
The prequantum Hilbert space has a  connection $\Theta$ with
its curvature equal to $-{i \over \hbar}\omega$.
In terms of the  coordinate system $(Z , Z^{\dagger})$ we can
write down the covariant derivative along a vector
field $V$, acting on a section of this
line bundle as,
\beq
     \nabla_V \psi=V^Z\partial_Z\psi + V^{Z^{\dagger}}\partial_{Z^\dagger}
                    \psi + i_V\Theta\psi
.\eeq
Here, $V^Z\partial_Z\psi=\lim_{t \to
0}{\psi(Z+tV^Z,Z^{\dagger})-\psi(Z,Z^{\dagger}) \over t}$.
We see that it is enough to choose the connection 1-form $\Theta$ such that
$d\Theta=-{i \over \hbar} \omega$. 
In the above coordinate system, we can express the symplectic form as
 $\omega=-2i\partial_Z\partial_{Z^\dagger}
\log {\rm det}(1_{(M-m)\times (M-m)}-Z^{\dagger}Z)$.
The choice of $\Theta$ is not unique, but a convenient one is,
\beq
    \Theta={1 \over \hbar}
(\Tr (1-Z^{\dagger}Z)^{-1}dZ^{\dagger}Z-\Tr (1-Z^{\dagger}Z)^{-1}Z^{\dagger}dZ)
\eeq
The coefficient of proportionality can be fixed
by comparing the value of $\omega$ with its expression in terms
of $\Phi$ at the point
$\Phi=\epsilon$, or $Z=0$.

Since $\omega$ is non-degenerate its $m(M-m)$th power
defines a volume form. ${\cal H}_{Pre}$ is given by square integrable
functions on the Disc with respect to this volume form.
The quantum Hilbert space is given by the holomorphicity requirement;
\beq
    {\cal H}_Q=\{ \psi | \nabla_{Z^{\dagger}} \psi=0  \quad \psi \in
    {\cal H}_{Pre} \}
.\eeq
It is possible to give a more explicit description  of  the holomorphicity
condition;
any function $\psi={\rm det}^{1 \over \hbar}(1-Z^{\dagger}Z)\Psi(Z)$
, where $\Psi(Z)$ is an ordinary holomorphic function of the variables $Z$,
belongs to ${\cal H}_Q$.
We will find the operators which correspond to the moment maps. Since
they generate the symmetry, they are a natural set of operators in
a quantization problem. We will use the corresponding prequantum operators
and obtain an operator acting on ${\cal H}_Q$.
Suppose that $f_{-u}$,
where this time $\epsilon u^{\dagger} \epsilon=-u$,
 is the moment map and $V_u=V^Z_u\partial_Z+V^{Z^{\dagger}}
_u\partial_{Z^{\dagger}}$ is the vector field generated
by it. Then we can calculate $\tilde f_{-u}$ action on
$\psi={\rm det}^{1 \over \hbar}(1-Z^{\dagger}Z)\Psi(Z)$;
\beq
     \tilde f_{-u}\psi=-i\hbar({\cal L}_{V^{Z}_u}+\Theta(V^{Z}))\psi+f_{-u}\psi
\eeq
where the antiholomorphic part dropped out.
We  note that the action of $V_u$ on $Z$ is given by;
${\cal L}_{V_u} Z=\alpha Z+\beta-Z\gamma Z-Z\delta$ where,
\beq
       u=\pmatrix{\alpha & \beta\cr
                  \gamma & \delta\cr}
\nn\eeq
is the decomposition of $u$ into block form; $\alpha^{\dagger}=-\alpha$
, $\beta^{\dagger}=\gamma$ and $\delta^{\dagger}=-\delta$.
We can explicitly write them down;
\bea
    \tilde f_{-u}\psi=&-&i\hbar[-{2 \over \hbar}\Tr(1-Z^{\dagger}Z)^{-1}
     Z^{\dagger}(\alpha Z+\beta-Z\gamma Z-Z\delta)]\psi\nn\\
         &+&i[\Tr\alpha(1-2(1-ZZ^{\dagger})^{-1})-\Tr\beta Z^{\dagger}
             (1-ZZ^{\dagger})^{-1}+\Tr \gamma(1-ZZ^{\dagger})^{-1}Z\nn\\
           &+&\Tr\delta (1+Z^{\dagger}(1-ZZ^{\dagger})^{-1}Z)]\psi
   -i\hbar {\rm det}^{1 \over \hbar}(1-Z^{\dagger}Z){\cal L}_{V_u}\Psi(Z).
\eea
A straightforward calculation will reveal that,
\beq
    \tilde f_{-u}\psi(Z,Z^{\dagger})=
   {\rm det}^{1 \over \hbar}(1-Z^{\dagger}Z)(-i\hbar)[{\cal L}_{V_u}\Psi(Z)
   +(-{2 \over \hbar}\Tr(\gamma Z + \delta) +{1 \over
\hbar}\Tr(\alpha+\delta))
   \Psi]
.\eeq
This shows that the action of the moment maps in fact preserves the
holomorphicity requirement, since all the terms inside the square brackets are
holomorphic functions of $Z$.
We can, therefore, define an action of the
Lie algebra $\underline U(m,M-m)$ on the space of holomorphic functions
generated by $\tilde f_{-u}$ by removing the determinant part;
\beq
   \tilde f_{-u}\Psi=-i\hbar[{\cal L}_{V_u}\Psi +(-{2 \over \hbar}
       \Tr (\gamma Z+\delta)+{1 \over \hbar}\Tr (\alpha +\delta))\Psi]
.\eeq
This
can be exponentiated to a group action of $U(m,M-m)$ if ${2 \over \hbar}$ is a
positive integer, $N_c$.
We can write down the group action explicitly;
\beq
     \rho(g^{-1})\Psi={\rm det}^{-{2 \over \hbar}}(cZ+d)
                {\rm det}^{1 \over \hbar}(g)\Psi\big((aZ+b)(cZ+d)^{-1}\big)
.\eeq
We see that it is a unitary representation of the symmetry group
$U(m,M-m)$ in the Quantum Hilbert space, i.e.
\bea
    <\rho(g^{-1})\Psi_1\!\!&,&\!\! \rho(g^{-1})\Psi_2>=\nn\\
         &=&\int  {\rm det}^{1 \over \hbar}(1-Z^{\dagger}Z)
        ({\rm det}^{1 \over \hbar}(1-Z^{\dagger}Z))^*(\rho(g^{-1})\Psi_1)^*
        \rho(g^{-1})\Psi_2 \omega^{m(M-m)}\nn\\
      &=&\int |{\rm det}^{1 \over \hbar}(1-Z^{\dagger}Z)|^2\Psi_1^*\Psi_2
        \ \omega^{m(M-m)}. \nn
\eea
by using the invariance of $\omega$ and the transformation property of
${\rm det}^{1 \over \hbar}(1-Z^{\dagger}Z)$ under a group action.
(The representation we found can be compared with the known 
series of representations in the literature about $SU(m,M-m)$ \cite{knapp})
The projection $K$, discussed earlier, is not necessary
in the case of moment maps, i.e.
$\hat f_{u}=K\tilde f_uK=\tilde f_uK$.

Let us identify the heighest weight corresponding to this representation. 
It is easiest to work with the Lie algebra. We consider 
the complexification of $\underline{U}(m,M-m)$: $\underline{GL}(M,{\bf C})$.
The heighest weight vector is given by $\Psi_0(Z)=1$. We can act with 
the upper triangular matrices and see that,
\beq
     \hat f_{-u^C}\Psi_0=[-\frac{1}{\hbar}\Tr \delta +\frac{1}{\hbar} \Tr 
      \alpha]\Psi_0
\eeq
 We recall that the heighest weight 
is given by the coefficients of the diagonal elements in the action of 
an upper triangular matrix\cite{humphreys,jacobson}. More explicitly,
$\rho(u^C)\Psi_0=\sum_i w_i u^C_{ii}\Psi_0$ gives the 
weights $w_i$. 
From the above expression we get
\beq 
      w_i=-\frac{N_c}{2} \ \ {\rm for} \ \ i\le m \ \ \ { \rm and}\ \ \
       w_i=\frac{N_c}{2} \ \ {\rm for} \ \ i>m 
.\eeq 
Since there is a unique highest weight, this representation of 
$U(m,M-m)$ is also
irreducible.  We can see further that this representation 
is equivalent to the algebraic one since they have the same weights.

One can check that given an arbitrary function $F(\Phi)$ we have
a vector field generated by $F$, $V_F=i[\Phi,{\partial F \over  \partial
\Phi}]$. If we obtain the prequantum  operator $\tilde F$ and act on
functions in ${\cal H}_Q$, we see that it gives,
\beq
    \tilde F\psi={\rm det}^{1 \over \hbar}(1-Z^{\dagger}Z)
  [(-2i\Tr {(1-ZZ^{\dagger})^{-1}
          Z^{\dagger}V^Z_F(Z,Z^{\dagger})}+F(Z,Z^{\dagger})-i\hbar
           {\cal L}_{V^Z_F}] \Psi(Z)
.\eeq
The expression  inside the square brackets is not in general holomorphic; thus
the prequantum operators do not preserve holomorphicity when
they act  on ${\cal H}_Q$.
We need to project back to the quantum Hilbert space, this will
be the quantum operator.
However one can give another rule
for quantization; since $f_u$'s are linear functions in $\Phi$ any
polynomial expression in $\Phi$ can be quantized by giving the expression
for $\Phi$. This rule preserves the holomorphicity and it is a well-defined
quantization process except for some  ordering ambiguities. But these are
higher order terms which are inherent in any quantization program. In our own
examples, normal ordering will be the most convenient.

Now, we will construct the quantum Hilbert space for $Gr_M(m)$.
Since there is no global coordinate system over the Grassmannian we
need to employ a different method.
The fact that the Grassmannian is a principal fiber bundle will be used to
construct a prequantum line bundle on it.
We  view the Grassmannian as a coset space
$U(M)/ U(m) \times U(M-m)$ and  take $U(m) \times U(M-m)$
as the group acting on $U(M)$. We
denote this  as $U(m) \times U(M-m) \to U(M) \to Gr_m(M)$.
Given  a one dimensional representation
of the group $U(m)\times U(M-m)$ we can construct a
line bundle. The most general
one dimensional representation of $U(m) \times U(M-m)$ on ${\bf C}$
is,
\beq
      \rho(h)=({\rm det}(h_1h_2))^{w} ({\rm det}(h_1))^{N_c}
\eeq
where we denote a typical element of the group by
 $\pmatrix{ h_1 & 0\cr 0 & h_2\cr}$ and $N_c$ and $w$ are integers.
Let the line bundle  defined by this representation be $L_\rho$:
\beq
	L_\rho=\big(U(M)\times_\rho {\bf C})/U(m)\times U(M-m)\big).
\eeq
 Sections of  $L_\rho$ are functions $\psi:U(M) \to {\bf C}$ satisfying
the equivariance condition,
\beq
     \psi(gh)=\rho(h)\psi(g).
\eeq
The combination $\psi_1^*(g)\psi_2(g)$ is  a function on the Grassmannian;
thus  we can define an inner product on the sections of $L_\rho$, by
using  the Liouville measure:
\beq
	<\psi_1,\psi_2>=\int_{Gr_m(M)}\psi_1^*\psi_2\omega^{m(M-m)}
\eeq
This defines the pre--quantum Hilbert space.
Given a connection on the principal fiber bundle,
 the covariant derivatives acting on the sections  are given by
the Lie derivatives along the horizontal directions, namely
\beq
     \nabla_X \psi = {\cal L}_{X^H} \psi
\eeq
where we use the fact that sections are given as equivariant functions
\cite{kobayashi}.
The principal bundle  $U(m)\times U(M-m) \to U(M)\to Gr_m(M)$ has a
connection induced from an invariant metric on $U(M)$: a horizontal vector is
 one that is orthogonal to all vertical vectors.

We can think of a vector field $X$ on $U(M)$ in terms of a matrix--valued
 function
on $U(M)$, and  describe the components of $X$ in the basis of right
invariant
 vector fields, which generates the left action.
 We can write the inner product of two vector fields $X_1,X_2$ is, in terms of
 components, using the standart invariant metric:
\beq
	<X_1,X_2>(g)=\tr X_1^{\dag}(g)X_2(g).
\eeq
The components of the vertical vector field $v_\xi$ corresonding to the Lie
 algebra element $\xi=\pmatrix{\xi_1&0\cr 0& \xi_2}$ is
\beq
	v_\xi(g)=ig\pmatrix{\xi_1&0\cr 0& \xi_2} g^{-1}.
\eeq
Thus a horizontal vector field  $h$ must  have components $h(g)$ satisfying
\beq
	\tr h(g)^{\dag}v_{\xi}(g)=0\hbox{ \rm for  all}\; \xi_1,\xi_2;
\eeq
equivalently,
\beq
	[\eps, gh(g)g^\dagger]_+=0.
\eeq
Given an arbitrary vector field $X$, the components of its vertical part $X^V$
 and horizontal part $X^H$ can now be seen to be
\beq
	X^H(g)={i\over 4}g[\eps,[\eps,g^{-1}X(g)g]]g^{-1},\quad
	X^V(g)={i\over 4}g[\eps,[\eps,g^{-1}X(g)g]_+]_+g^{-1}
\eeq
This defines a connection on the Principal bundle. Equivalently,
the connection is given by a Lie algebra valued one form $\Omega$ on
the total space, such that 
\beq
     \Omega(V_\xi)=\xi, \quad \Omega(X^H)=0
\eeq
for $V_\xi$ generated by $\xi=\pmatrix{\xi_1&0\cr 0 & \xi_2\cr}$
written by using the right action.

We will now derive a formula for the covariant derivative of
 a section of $L_\rho$. Let $V(\Phi)=i[U(\Phi),\Phi]$ be the components of a
 vector field on the Grassmannian. Using the identification $\Phi=g\eps
g^{-1}$,
we can think of this as a vector field on $U(M)$;
  	$X(g)=U(g\eps g^{-1})$.
The Lie derivative along the vertical part of $X$ is
\bea
	{\cal L}_{X^V}\psi(g)&=&{d\over dt}\psi\big([1+tX^V(g)]g\big)_{t=0}
		={d\over dt}\psi\big(g[1+tg^{-1}X^V(g)g]\big)_{t=0}\nn\\
  &=&\underline \rho\big( {i\over 4}[\eps,[\eps,g^{-1}X(g)g]_+]_+\big)\psi,\nn
\eea
where in
 the last step we have used the infinitesimal version of the equivariance
 condition on $\psi$ and use $\underline \rho$
to denote the representation of the
Lie algebra.
The covariant derivative along the vector field $V$ on the Grassmannian is now,
\beq
	\nabla_X \psi(g)={\cal L}_X \psi(g)-{\cal L}_{X^V}\psi(g)
                        =({\cal L}_X-\underline\rho(\Omega(X)))\psi(g)
\eeq
For our line bundle we can write this down as,
\beq
    \nabla_X\psi(g)={\cal L}_X\psi(g)-i[w\Tr X(g)+
                    {N_c \over 2}\Tr(1-\Phi)X(g)]\psi(g)
\eeq
We can then calculate the curvature,
\beq
\big([\nabla_X,\nabla_Y]-\nabla_{[X,Y]}\big)\psi(g)={-iN_c\over
 2}\omega(X,Y)\psi(g).
\eeq

 We identify the $L_\rho$ with the prequantum line bundle. Because
the curvature of the covariant derivative  on this line bundle
must be $-{i \over \hbar}\omega$, we see that in this theory $\hbar$
can only take discrete values given by $2 \over N_c$.

Of particular interest are the moment maps,
$f_u(\Phi)=-\Tr u\Phi$,
which generate the group action:
$V_{f_u}=i[u,\Phi]$.
The pre--quantum operator corresponding to $f_u$ is just,
\beq
	\tilde f_u\psi(g)=-i\hbar L_u\psi(g)
               -[\hbar w+1]\Tr u\psi(g). \label{q1.4}
\eeq
where	$L_u \psi(g)={d \over dt}\psi([1+itu]g)|_{t=0}$.
These operators as expected, provide  a representation of the Lie algebra
 ${\underline U}(M)$.

The pre--quantum operators provide a highly reducible representation of the
 Poisson algebra of functions on the Grassmannian. To obtain an irreducible
 representation, we use  the complex structure on the Grassmannian.
Since $\omega$ is of type $(1,1)$, at least locally there are solutions to the
 holomorphicity condition,
\beq
	\nabla_X \psi=0 \quad \hbox{\rm for $X$ type (0,1)}.
\eeq
We will now construct global  solutions to this conditions by viewing the
 Grassmannian as a homogeneous space of $GL(M,{\bf C})$. In fact
\beq
	B_m\to GL(M,{\bf C})\to Gr_m(M)
\eeq
is a holomorphic principal bundle.  Our line bundle $L_\rho$ can be given a
 holomorphic structure by extending  $\rho$ to
 a holomorphic representation of $B_m$:
\beq
   \rho\big(\pmatrix{a&b\cr 0&d}\big)= ({\rm det}(ad))^{w}({\rm det} a)^{N_c}
\eeq
for integer $w$  and $N_c$.
Thus the holomorphic sections of the line bundle are given by
holomorphic functions $\psi:GL(M,{\bf C})\to {\bf C}$
satisfying the equivariance condition:
\beq
	\psi\big(\gamma\pmatrix{a&b\cr 0&d}\big)
          =({\rm det}(ad))^{w}({\rm det} a)^{N_c}\psi(\gamma).
\eeq
Some solutions to this condition can be found as follows:
\beq
      \psi(\gamma)=({\rm det} \gamma)^w ({\rm det} \gamma_-)^{N_c}
\eeq
where $\gamma_-$ is any $m\times m$ submatrix of the first column (i.e.,
 $\pmatrix{\gamma_{11}\cr \gamma_{21}}$) of $\gamma$:
\beq
     \pmatrix{ (\gamma_{11})_{m \times m} & (\gamma_{12})_{M-m\times m}\cr
               (\gamma_{21})_{M-m \times m} & (\gamma_{22})_{M-m \times
M-m}\cr}
.\eeq
More generally, we can write down several different submatrices of the
the same type, $\gamma_{1-},\gamma_{2-},...\gamma_{p-}$ and write down
the general solution as
\beq
      \psi(\gamma)=({\rm det} \gamma)^w({\rm det} \gamma_{1-})^{k_1}
          ({\rm det} \gamma_{2-})^{k_2}....({\rm det} \gamma_{p-})^{k_p}
.\eeq
where $k_1+k_2+...+k_p=N_c$. For such global solutions to the polarization
condition to exist, each $k_1,...,k_p$ must be a positive integer;
 thus $N_c$ is  a positive integer. The above solutions are not all
linearly independent, but one can find a linearly independent set among them.
 If we restrict to the
 subgroup $U(M)$, these are  equivariant under $U(m)\times U(M-m)$.
 Moreover, they are also square integrable with respect  to the inner
 product defined in the previous section. Their linear span  defines the
physical Hilbert
 space ${\cal H}_Q$. This is a subspace of ${\cal H}_{Pre}$, consisting of
those equivariant
 functions on $U(M)$ that have a holomorphic extension to $GL(M,{\bf C})$.
We note that {\it global} holomorphic sections exist for positive integer
values of $N_c$, which in turn fixes the value of $\hbar$.

There is a representation $r$ of the group $GL(M,{\bf C})$ on ${\cal H}_Q$:
\beq
	r(\gamma)\psi(\gamma')=\psi(\gamma^{-1}\gamma').
\eeq
We will now see that this is an irreducible  representation;
 moreover we will identify it with a particular tensor space.
A highest weight vector in ${\cal H}_Q$ is defined by the condition
\beq
	\psi(\lambda\gamma)=\prod_i\lambda_{ii}^{w_i}\psi(\gamma)
\eeq
for lower triangular matrices $\lambda$. The infinitesimal version of this
 condition gives the familiar highest weight condition on the Lie algebra
 representation. The weight vector is $w_i$. Now, we will show that in 
${\cal H}_Q$
 there is  ( up to constant multiple) one such vector. For  $\gamma$ in an
 open dense subset of $GL(M,{\bf C})$ we can write
\beq
	\gamma=\lambda\tau,
\eeq
where $\lambda$ is lower triangular and $\tau$ is upper triangular. If the
 equivariance and highest weight conditions are  both satisfied, the function
$\psi$ is determined in this open subset up to a constant:
\beq
	\psi(\gamma)=\prod_i\lambda_{ii}^{w_i}\rho(\tau)\psi(1).
\eeq
This is because upper triangular matrices are automatically in $B_m$. But then
 $\psi$ has at most one  holomorphic extension to $GL(M,{\bf C})$.
One such  vector in ${\cal H}_Q$ can be explicitly displayed;
it is a special
 case of the previous example
\beq
	\psi(\gamma)=[\det \gamma]^w[\det \gamma_{11}]^{N_c}.
\eeq
Thus there is exactly one highest weight vector in our representation of
$GL(M,{\bf C})$ on the quantum Hilbert space.
Moreover, we can determine the  weight $w_i$ as well:
\beq
	w_i=w+N_c\quad\hbox{\rm for}\;
      i=1\cdots m,\quad w_i=w\;\hbox{\rm for}\ i>m.
\eeq

We  note that the action of moment maps, given by
 the formula (\ref{q1.4}),  on these sections preserve the
holomorphicity condition. From their expression we see
that they  act as Lie derivatives; the operators corresponding to
 the moment maps, provide a
 representation of the  Lie algebra $\underline U(M)$
the  holomorphic sections of $L_\rho$. Since there is exactly one highest
weight vector, this representation is also irreducible.
If we compare this to the representation we obtain on the fermionic Fock space
we see that they have the same highest weight vector.  Thus our approach by
geometric quantization produces the same representation as the algebraic
quantization on the fermionic Fock space.

Now let us  consider the more general  case of a Hamiltonian vector field of an
arbitrary function $F(\Phi)$: $V_F(\Phi)=i[\Phi,{\pdr F\over \pdr \Phi}]$
and $ U_F(\Phi)={\pdr F\over \pdr \Phi}$.
We can combine this with the previous results to get a formula for the
 pre--quantum operator corresponding to $F$:
\beq
   \tilde F\psi(g)=-\hbar\Tr {\pdr F\over \pdr \Phi}g
       {\pdr \psi \over \pdr g}
+[\hbar w\Tr {\pdr F\over \pdr \Phi}+\Tr (1-\Phi){\pdr F\over \pdr
 \Phi}]\psi(g)+F(\Phi)\psi(g).
\eeq
We see that, in general, the above expression cannot preserve holomorphicity;
it involves an explicit $\Phi$ dependence unless $F$ is linear in $\Phi$.
$\Phi$ is not holomorphic since there are no non-constant
global holomorphic functions on
a compact manifold.
In general, we need to  project to the holomorphic subspace.
But this is a complicated procedure. We are mostly interested in
polynomial functions of $\Phi$. In this case
there is a simpler  alternative. In a polynomial expression
 we can replace the matrix elements
of each $\Phi$ by the
operator corresponding to it through the moment maps. This procedure
is ambiguous, because the operators corresponding to the different matrix
elements of $\Phi$ do not commute. However, these ambiguities can be resolved
by some ordering rule, such as symmetric ordering.
Since moment maps preserve the holomorphicity,
 operators constructed this way provide  a quantization of polynomial
functions in ${\cal H}_Q$.

\end{document}